\documentclass[journal,onecolumn]{IEEEtran}
\usepackage[margin=1in]{geometry}
\usepackage{amsmath} % for \eqref

\usepackage{amssymb}
\usepackage{multirow}
\usepackage{tabu}
\usepackage{mathtools}
\usepackage[compress]{cite}
\usepackage{subfigure}

\usepackage{url}
\usepackage{graphicx}
\usepackage[dvipsnames,table,xcdraw]{xcolor}
\usepackage{verbatim}% http://ctan.org/pkg/verbatim
\makeatletter
\newcommand{\verbatimfont}[1]{\def\verbatim@font{#1}}%
\makeatother
\verbatimfont{\ttfamily\small}

\newcommand{\bi}{\begin{itemize}}\newcommand{\ei}{\end{itemize}}
\newcommand{\be}{\begin{equation}}\newcommand{\ee}{\end{equation}}
\newcommand{\bee}{\begin{enumerate}}\newcommand{\eee}{\end{enumerate}}
\newcommand{\bea}{\begin{eqnarray}}\newcommand{\eea}{\end{eqnarray}}
\newcommand{\beas}{\begin{eqnarray*}}\newcommand{\eeas}{\end{eqnarray*}}
\newcommand{\bc}{\begin{center}}\newcommand{\ec}{\end{center}}
\usepackage[left,pagewise]{lineno} 
\usepackage[english]{babel}

\newcommand{\sbt}{\,\begin{picture}(-1,1)(-1,-3)\circle*{3}\end{picture}\ }
\newcommand{\pr}{p}

\newcommand{\Thigh}{T_\uparrow}

\newcommand{\figsize}{0.35\textwidth}
\newcommand{\defeq}{\mathrel{:\mkern-0.25mu=}}
\newcommand{\set}{}

\DeclareMathOperator*{\erfc}{erfc} 
\DeclareMathOperator*{\argmin}{argmin} 
\DeclareMathOperator*{\argmax}{argmax} 
\DeclareMathOperator{\Expect}{{\rm I\kern-.3em E}}
\newcommand\E[1]{\Expect \left[{#1}\right]}

\title{Human Trust-based Feedback Control\\
\Large Dynamically varying automation transparency to optimize human-machine interactions}
\author{Kumar~Akash,
        Griffon~McMahon,
        Tahira~Reid,
        and~Neera~Jain% <-this % stops a space
\thanks{The authors are with the School of Mechanical Engineering, Purdue University, West Lafayette, IN, 47907 USA (e-mail: kakash@purdue.edu, gmcmahon@purdue.edu, neerajain@purdue.edu, tahira@purdue.edu). This work has been accepted for publication in the IEEE Control Systems Magazine}% <-this % stops a space
}
\newif\ifPDF \ifx\pdfoutput\undefined\PDFfalse \else\ifnum\pdfoutput > 0\PDFtrue \else\PDFfalse \fi \fi
\ifPDF 
\usepackage{hyperref}
\hypersetup{plainpages = false, colorlinks=true, linkcolor=black, citecolor = green!50!blue, urlcolor = blue, filecolor=black, hypertexnames=false}
\fi

\begin{document}
\IEEEpubid{\begin{minipage}{\textwidth}\ \\[12pt]
  \copyright~2020 IEEE. Personal use of this material is permitted.  Permission from IEEE must be obtained for all other uses, in any current or future media, including reprinting/republishing this material for advertising or promotional purposes, creating new collective works, for resale or redistribution to servers or lists, or reuse of any copyrighted component of this work in other works.
\end{minipage}} 

\maketitle

\begin{abstract}
Human trust in automation plays an essential role in interactions between humans and automation. While a lack of trust can lead to a human's disuse of automation, over-trust can result in a human trusting a faulty autonomous system which could have negative consequences for the human. Therefore, human trust should be \emph{calibrated} to optimize human-machine interactions with respect to context-specific performance objectives. In this article, we present a probabilistic framework to model and calibrate a human's trust and workload dynamics during his/her interaction with an intelligent decision-aid system. This calibration is achieved by varying the automation's transparency---the amount and utility of information provided to the human. The parameterization of the model is conducted using behavioral data collected through human-subject experiments, and three feedback control policies are experimentally validated and compared against a non-adaptive decision-aid system. The results show that human-automation team performance can be optimized when the transparency is dynamically updated based on the proposed control policy. This framework is a first step toward widespread design and implementation of real-time adaptive automation for use in human-machine interactions. 
\end{abstract}

\section{Introduction}

% Motivation
Automation has become prevalent in the everyday lives of humans. However, despite significant technological advancements, human supervision and intervention are still necessary in almost all sectors of automation, ranging from manufacturing and transportation to disaster-management and healthcare~\cite{wang2017trends}. Therefore, we expect that the future will be built around \emph{human-agent collectives}~\cite{jennings2014humanagent} that will require efficient and successful interaction and coordination between humans and machines. It is well established that to achieve this coordination, human trust in automation plays a central role~\cite{muir1987trust,lee2004trust,sheridan2005humanautomation}. For example, the benefits of automation are lost when humans override automation due to a fundamental lack of trust~\cite{muir1987trust,sheridan2005humanautomation}, and accidents may occur due to human mistrust in such systems~\cite{richtel2015googles}. Therefore, trust should be appropriately \emph{calibrated} to avoid disuse or misuse of automation~\cite{lee2004trust}.

% Problem Definition
These negative effects can be overcome by designing autonomous systems that can adapt to a human's trust level. One way to adapt automation based on human trust is to augment the user interface with more information---either raw data or processed information---to help the human make an informed decision. This ``amount of information'' has been defined as automation \emph{transparency} in the literature. Transparency has been defined as ``the descriptive quality of an interface pertaining to its abilities to afford an operator's comprehension about an intelligent agent's intent, performance, future plans, and reasoning process'' \cite{chen2014situation}. With higher levels of transparency, humans have access to more information to aid their decisions, which has been shown to increase trust \cite{mercado2016intelligent, helldin2014transparency}. Therefore, adaptive automation can be implemented by varying automation transparency based on human trust. However, because high transparency requires communicating more information, it can also increase the workload of the human \cite{lyu2017drivers}. In turn, high levels of workload can lead to fatigue and therefore reduce the human's performance. Therefore, we aim to design adaptive automation that can vary automation transparency to accommodate changes in human trust and workload in real-time to achieve optimal or near-optimal performance. This requires a dynamic model of human trust-workload behavior to design and implement control policies. 

% Gaps in literature
In existing literature, some of these human behaviors have been studied and modeled. Researchers have developed various models of human trust \cite{lee1992trust,moe2008learning,malik2009web,xu2015optimo} and have also shown that transparency has a significant effect on trust \cite{helldin2014transparency,mercado2016intelligent} as well as workload \cite{bohua2011drivers,lyu2017drivers}. Nevertheless, there is no framework that captures the \emph{dynamic} effect of automation transparency on human trust-workload behavior. Furthermore, a fundamental gap remains in using such a framework to develop control policies for influencing human behavior through changes in transparency to improve human-machine interaction. 

\begin{figure}
  \centering
  \includegraphics[width=.55\textwidth]{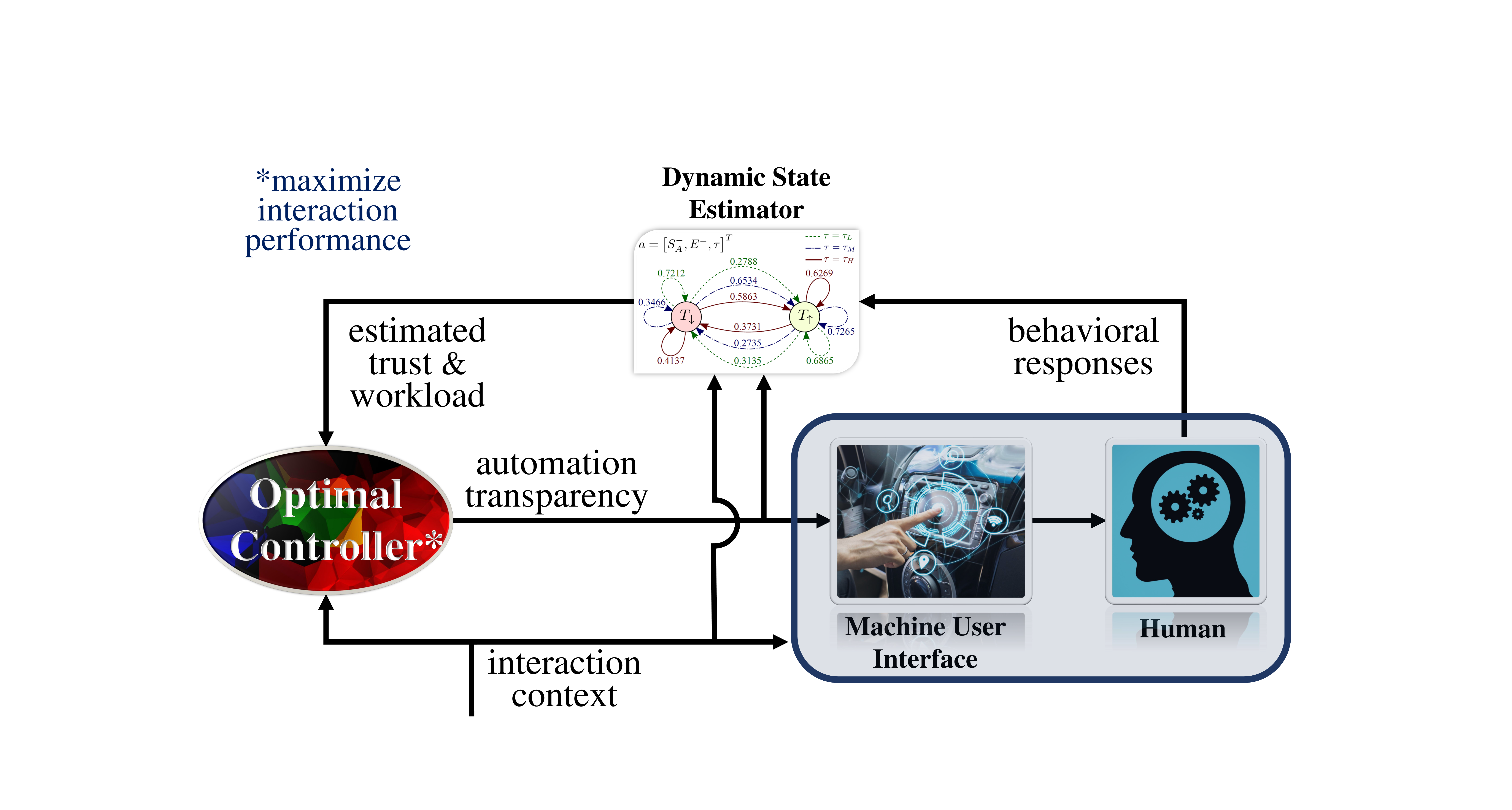}
  \caption{Block diagram depicting a trust and workload-based feedback control architecture for optimizing human-machine interactions. The human behavior model is used to estimate the non-observable human states of trust and workload using the machine outputs, the interaction context, and the observable human responses. An optimal control policy dynamically varies automation transparency based on the estimated human states to maximize a context specific performance objective.}
  \label{fig_block_diagram_intro}
\end{figure}

In this article, we present a trust-based feedback control framework, shown in Figure~\ref{fig_block_diagram_intro}, that improves context-specific performance objectives using automation transparency during human-machine interactions. We show that human trust and workload can be modeled and predicted based on a human's behavioral responses to the automation's decision-aids and can subsequently be used as feedback to vary automation transparency. Preliminary results of the work presented in this article have been published in earlier work \cite{akash2018improving,akash2018improving-1}. This work significantly expands on prior work with a re-designed interface for the experiment, an improved modeling framework, and a comprehensive experimental validation of the control policy. The scope of this article is restricted to decision-aid systems, but extensions to other types of automation are briefly discussed. 
\section{Levels of Automation and Transparency}

\begin{figure}
  \centering
  \includegraphics[width=0.35\textwidth]{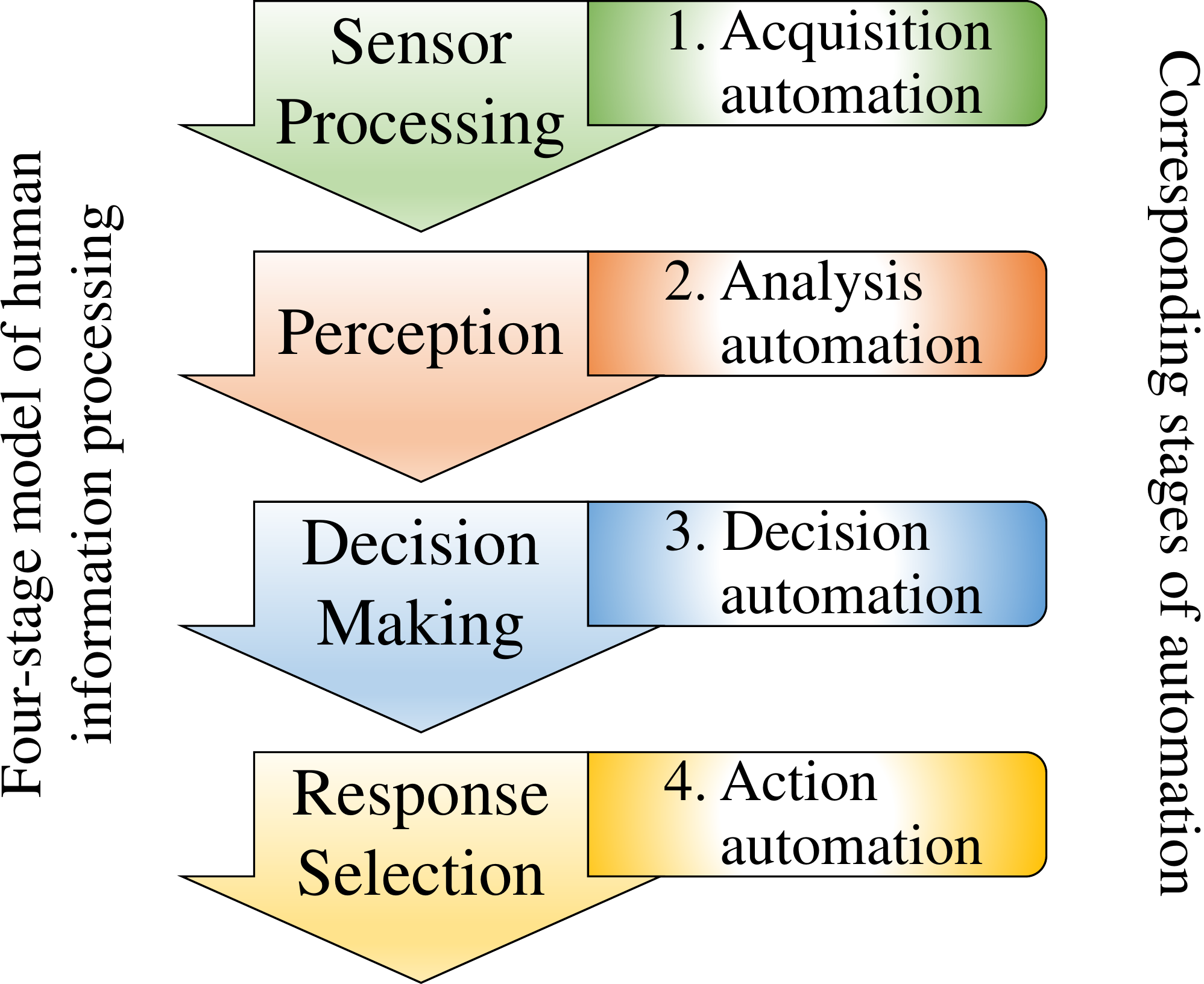}
  \caption{Simple four-stage model of human information processing and the corresponding types of automation (adapted from \cite{parasuraman2000model}). Each stage of human information processing has its equivalent in system functions that can be automated, thereby leading to four types of automation.}
  \label{fig:intro_LOAs}
\end{figure}

Parasuraman et al. proposed that most systems involve  four stages of sequential tasks with each successive stage dependent on successful completion of the previous one. They correspond to the four information-processing stages of humans: (1) information acquisition (sensory processing), (2) information analysis (perception), (3)  decision and action selection (decision-making), and (4) action implementation (response selection) \cite{parasuraman2000model,parasuraman2008humans}. This four-stage model of human information processing has its equivalence in system functions that can be automated, leading to four types of automation as shown in Figure~\ref{fig:intro_LOAs}.
These can be conveniently called acquisition, analysis, decision, and action automation, respectively. Within this four-stage model, each type of automation can have different degrees or levels of automation (LOA), depending on the context. 

Information sources are inherently uncertain due to factors including sensor imprecision and unpredictable events. Nevertheless, even imperfect acquisition and analysis automation at a reliability as low as $70\%$, can improve performance as compared to unaided human performance \cite{wickens2007benefits}. On the other hand, the utility of \emph{decision} and \emph{action} automation is more sensitive to adequately calibrated human trust in the automation than \emph{acquisition} and \emph{analysis} automation.  This is because in the case of decision automation, the human is typically being asked to comply, or not comply, with the decision proposed by the automation. Similarly, in the case of action automation, the human is being asked to rely, or not rely, on the actions being taken by the automation.  In both cases, how strongly the human \emph{trusts} the automation will affect their compliance with the automation (or the lack thereof).  This can be dangerous in cases when the human's trust is not adequately calibrated to the reliability of the automation, such as when a human trusts an automation's faulty decision.  It is possible to mitigate this issue by varying the LOA depending on situational demands during operational use; this has been classically defined as adaptive automation \cite{hancock1985adaptive,rouse1988adaptive,parasuraman1992theory}. For decision-aid systems, adaptive automation can be realized by adjusting, or controlling, the amount of raw or processed information given to the human or in other words, controlling \emph{transparency.} See~\cite{alonso2018system} for a review on system transparency during human automation interactions.

\section{Effects of Transparency on Trust and Workload}

Several researchers have investigated the effect of transparency on trust, where systems with higher transparency have been shown to enhance humans' trust in systems \cite{tintarev2007survey,felfernig2006empirical,sinha2002role}. Studies described in ~\cite{wang2016impact,wang2016trust,wang2015intelligent} showed that only the robot's reliability influenced trust; however, the studies also highlighted the limitation of the use of self-reported trust data. Notably, researchers have argued that high transparency can even reduce trust if the information provided is not interpretable or actionable to humans \cite{hosseini2018four}. Furthermore, too much detail communicated through higher transparency can increase the time required to process the information \cite{tintarev2007survey} and distract the human from critical details \cite{ananny2018seeing}. Researchers have shown that cognitive difficulty, and thereby cognitive workload, increases with an increase in information \cite{bohua2011drivers}. Findings in \cite{helldin2014transparency} suggest that increased system transparency increases trust in the system but also causes workload to increase. Although a recent study based on the findings of \cite{helldin2014transparency} found that workload need not necessarily increase with greater transparency, several researchers agree that a trade-off between increased trust and increased workload exists when considering increased transparency \cite{alonso2018system}. Therefore, to vary the automation transparency, we propose to first model both human trust as well as human workload in a quantitative framework.

\section{Modeling Human Trust and Workload} \label{sec_model}

Researchers have developed various models for human trust. Qualitative models \cite{moray1999laboratory,muir1994trust,jian2000foundations, desai2012modeling} are useful for defining which variables affect trust but are insufficient for making quantitative  predictions. On the other hand, regression models \cite{devries2003effects, muir1996trust} quantitatively capture trust but do not consider its dynamic response characteristics. To fill this gap, researchers have proposed both deterministic models  \cite{lee1992trust,lee1994trust,moray2000adaptive,lee2004trust,lee2006extending,akash2017dynamic,hu2018computational,hoogendoorn2013modelling} and probabilistic models \cite{cohen1998trust,moray1999laboratory,xu2015optimo} of human trust dynamics. With respect to probabilistic approaches, several researchers have modeled human trust behavior using Markov models, particularly hidden Markov models (HMMs) \cite{moe2008learning,malik2009web,elsalamouny2009hmmbased}. 
While HMMs can be used for intent inference and to incorporate human behavior related uncertainty\cite{li2003recognition,pineau2003pointbased,wang2009hmm,liu2012modeling}, they do not include the effects of inputs or actions from autonomous systems that affect human behavior. On the other hand, models based on Markov decision processes (MDP) do consider the effect of inputs or actions and have been used to model human behavior for human-in-the-loop control synthesis \cite{feng2016synthesis}. However, MDPs do not account for the unobserved nature of human cognitive constructs like trust and associated uncertainties. A useful extension of HMMs and MDPs, called partially observable Markov decision processes (POMDPs), provides a framework that accounts for actions/inputs as well as unobserved states and also facilitates calculating the optimal series of actions based on a desired reward function. Recent work has demonstrated the use of a POMDP model with human trust dynamics to improve human-robot performance~\cite{chen2018planning}. POMDPs have also been used in HMI for automatically generating robot explanations to improve performance \cite{wang2016impact, wang2016trust, wang2015intelligent} and estimating trust in agent-agent interactions \cite{seymour2009trustbased}. For example, the POMDP model in \cite{wang2016impact, wang2016trust, wang2015intelligent} is used to simulate only the dynamics of the robot's decisions and generates recommendations of different transparency levels. However, the model does not capture human trust-workload behavior nor the dynamic effects of automation transparency on that behavior. In this work, we model the human trust-workload behavior as a POMDP and optimally vary automation transparency to improve human-machine interactions.

Trust and workload levels of humans have been classically obtained using self-reported surveys. Trust surveys involve questions customized to an experiment along with a Likert scale for the participants to report how much they trusted the system and understood the scenario~\cite{jian2000foundations}. Workload is commonly assessed using the NASA TLX survey \cite{proctor2018human}.
In the context of real-time feedback algorithms, however, it is not practical to use surveys for human measurements because continuously inquiring humans is generally not feasible. Alternatively, we can use behavioral metrics that are readily available in real time and correlate to human trust-workload behavior, including compliance and response time. Compliance is defined as the human agreeing to the automation's recommendation when one is issued. Human response time is the time a human takes to respond to a stimulus \cite{luce1986response}. These metrics can be implicitly used to infer the underlying trust and workload states of the human. 

Several studies have shown a strong correlation between human trust and compliance. For example, researchers have shown that perception of trust is associated with improved compliance \cite{braithwaite1994trust}. Furthermore, studies showed that trust and compliance exhibited similar patterns with variations in system accuracy \cite{fox1996effects,fox1998effects}. Studies in \cite{wang2016trust} also confirmed the correlation between trust and compliance during human-robot interaction.  Similarly, other studies have shown a correlation between workload and response time\cite{helldin2014transparency,makishita2008differences}. The peripheral detection task (PDT) method based on human response time  has been shown to be a sensitive measure of cognitive workload \cite{crundall1999peripheral,patten2004using}. Newell and Mansfield showed that with environmental stressors, as participants’ reaction times slowed down, simultaneously their workload demands based on the NASA TLX assessment also increased \cite{newell2008evaluation}. Therefore, in this work, we assume a causal relationship of trust affecting compliance and propose to use response times as observations corresponding to workload.

\subsection{POMDP Model of Human Trust and Workload}

Here we consider contexts that involve human interaction with a decision-aid system that gives recommendations based on the presence or absence of a stimulus. During such an interaction, the final decision and action is taken by the human; the decision-aid system only provides a \emph{recommendation} to the human. Although such systems are a subset of autonomous systems, they are widely used in safety-critical situations, such as assistive robots used for threat detection in the military theater or health recommender systems used for detecting diseases. When interacting with a decision-aid system, the human is typically choosing to either comply with, or reject, the system's recommendation. This human decision has an associated response time ($RT$). Since we cannot directly observe human trust and workload states, we use human observations---compliance and response time---to estimate the states.  We further assume that human trust and workload are influenced by characteristics of the decision-aid system's recommendations. In particular, we consider the effects of the system's recommendation and transparency, and the human's past experience with the system. Also, the previous states of trust and workload affect the current state. Therefore, with an assumption that the dynamics of human trust and workload follow the Markov property \cite{puterman2014markov}, we use a POMDP to model the human trust-workload behavior. See Appendix~\ref{sb_POMDP} for more details on the structure of a POMDP and its similarities to state-space models.

We define the finite set of states $\mathcal{S}$ consisting of tuples containing the \textit{Trust} state $s_T$ and the \textit{Workload} state $s_W$, respectively, where %\change[NJ]{both can can be either low~($\sbt_\downarrow$) or high  ($\sbt_\uparrow$)}
each state can take on a low~($\sbt_\downarrow$) or high  ($\sbt_\uparrow$) value. The characteristics of the system recommendations are defined as the finite set of actions $\mathcal{A}$, consisting of tuples containing \textit{Recommendation},  \textit{Experience}, and \textit{Transparency}. Here, \textit{Recommendation} of the automation $a_{S_A}$ can be either Stimulus Absent $S_A^-$ or Stimulus Present $S_A^+$; \textit{Experience}~$a_E$, which depends on the reliability of the last recommendation, can be either Faulty $E^-$ or Reliable~$E^+$; and \textit{transparency}~$a_\tau$ can be either Low Transparency $\tau_L$, Medium Transparency $\tau_M$, or High Transparency $\tau_H$. The three levels of transparency depend on the context and the automation. The observable characteristics of the human's decision are defined as the set of observations $\mathcal{O}$ consisting of tuples containing \textit{compliance} and \textit{Response Time}. Here, \textit{Compliance} $o_C$ can be either Disagree $C^-$ or Agree $C^+$ and \textit{Response Time} $o_{RT} \in \mathbb{R}^+$ is defined as the time the human takes to respond after receiving the decision-aid's recommendation. The definition of our trust-workload POMDP model is summarized in Table~\ref{tab_model}.

\begin{table}
    \caption{Definition of the trust-workload POMDP model. Human trust and workload are modeled as hidden states that are affected by actions corresponding to the characteristics of the decision-aid's recommendations. The observable characteristics of the human's decisions are modeled as the observations of the POMDP.} \label{tab_model}
    \centering
{\tabulinesep=0.7mm
\begin{tabu}{|l|l|l|}
\hline
\multirow{2}{*}{
$\begin{matrix*}[l] 
\text{States} \\ s 
\in \mathcal{S}
\end{matrix*}$}       & \multirow{2}{*}{
            $s = \begin{bmatrix*}[l] 
            \text{\textit{Trust} } s_T\\ 
            \text{\textit{Workload} } s_W
            \end{bmatrix*}$} & 
                                    $s_T \in  \set{T} \defeq \begin{Bmatrix*}[l]  
                                    \text{Low Trust } T_\downarrow, \\ 
                                    \text{High Trust } T_\uparrow 
                                    \end{Bmatrix*}$ \\ \cline{3-3} 
& &                                 $s_W \in \set{W} \defeq \begin{Bmatrix*}[l]    
                                    \text{Low Workload } W_\downarrow, \\
                                    \text{High Workload } W_\uparrow 
                                    \end{Bmatrix*}$ \\ \hline
\multirow{3}{*}{
$\begin{matrix*}[l] 
\text{Actions} \\ a 
\in \mathcal{A}
\end{matrix*}$}       & \multirow{3}{*}{$
            a = \begin{bmatrix*}[l] 
            \text{\textit{Recommendation} } a_{S_A} \\ 
            \text{\textit{Experience} } a_E \\ 
            \text{\textit{Transparency} } a_\tau 
            \end{bmatrix*}$} & 
                                    $a_{S_A} \in  \set{S}_A \defeq \begin{Bmatrix*}[l]  
                                    \text{Stimulus Absent } S_A^-, \\
                                    \text{Stimulus Present } S_A^+ 
                                    \end{Bmatrix*}$ \\ \cline{3-3} 
& &                                 $a_E \in  \set{E} \defeq \begin{Bmatrix*}[l]  
                                    \text{Faulty last experience }E^-, \\
                                    \text{Reliable last experience }E^+ 
                                    \end{Bmatrix*}$ \\ \cline{3-3} 
& &                                 $a_\tau \in  \set{\tau} \defeq \begin{Bmatrix*}[l]  
                                    \text{Low Transparency }\tau_L, \\
                                    \text{Medium Transparency }\tau_M, \\
                                    \text{High Transparency }\tau_H 
                                    \end{Bmatrix*}$ \\ \hline
\multirow{2}{*}{
$\begin{matrix*}[l] 
\text{Observations} \\ o 
\in \mathcal{O}
\end{matrix*}$}       & \multirow{2}{*}{$
            o = \begin{bmatrix*}[l]
            \text{\textit{Compliance} } o_C \\
            \text{\textit{Response Time} } o_{RT} 
            \end{bmatrix*}$} & 
                                    $o_C \in  \set{C} \defeq \begin{Bmatrix*}[l] 
                                    \text{Disagree }C^-, \\
                                    \text{Agree }C^+ 
                                    \end{Bmatrix*}$ \\ \cline{3-3} 
& &                                 $o_{RT} \in \mathbb{R}^+$ \\ \hline
\end{tabu}}
\end{table}

We assume that human trust and workload behavior are conditionally independent given an action. Furthermore, we assume that trust only affects compliance, and workload only affects response time. This enables the trust and workload models to be identified independently. Moreover, it significantly reduces the number of parameters in each model and in turn, the amount of human data needed for training each model. The transition probability functions for the trust model, $\mathcal{T}_T:\set{T} \times \set{T} \times \mathcal{A}\rightarrow [0,1]$, and for the workload model, $\mathcal{T}_W:\set{W} \times \set{W} \times \mathcal{A}\rightarrow [0,1]$, are represented by $2\times2\times12$ matrices mapping the probability of transitioning between the states of $\text{trust } s_T\in  \set{T}$ or $\text{workload } s_W\in    \set{W}$ after an action $a\in\mathcal{A}$. For the trust model, the emission probability function $\mathcal{E}_T: \set{C} \times \set{T} \rightarrow [0,1]$ is represented by a $2\times 2$ matrix, mapping the probability of observing $\text{Compliance } o_C \in  \set{C}$ given the state of trust $s_T$. Similarly, for the workload model, the emission probability function $\mathcal{E}_W: \mathbb{R}^+ \times \set{W}\rightarrow [0,1]$ is represented by two probability density functions, each representing the probability of observing a response time $o_{RT} \in \mathbb{R}^+$ given the state of workload $s_W$. Human reaction-time has been shown to have a distribution similar to the ex-Gaussian distribution \cite{luce1986response,balota1999word,whelan2008effective}, which is a convolution (mixture) of a Gaussian and an exponential distribution. Here we assume that each workload state has a characteristic response time distribution defined by an ex-Gaussian distribution. See Appendix~\ref{sb_exGauss} for more details on the ex-Gaussian distribution and its use for modeling human response time.

\subsection{Human Subject Study} \label{sec_studydesign}

To parameterize the human trust and workload models, we collect human subject data in a specific decision-aid system context. The experiment presented here is adapted from an earlier study~\cite{wang2015intelligent}.  The modified experiment  captures the effects of different levels of system transparency on human trust and workload behavior along with the human-robot interaction performance. 

\emph{Stimuli and Procedure: } A within-subjects study was performed in which participants were asked to interact with a simulation consisting of multiple reconnaissance missions. Each participant performed three missions while being assisted by a decision-aid robot. In each mission, participants were required to search 15 buildings and mark them as safe or unsafe based on the presence or absence of gunmen. The goal of each mission was to search all of the buildings as fast as possible. Prior to entering each building, the participant needed to decide if they would use light armor or heavy armor while searching the building. They were informed that searching a building with heavy armor would take approximately 7 seconds but would ensure that they would not be injured if gunmen were present. On the other hand, searching with light armor would take only 3 seconds, but if gunmen were present, the participant would be injured and penalized with a 20-second recovery time. To assist the participant, the decision-aid robot would survey each building first and make a recommendation on which armor to use.

\begin{figure}
\centering
\subfigure[\label{fig_trans_low}]{\includegraphics[width=0.32\textwidth]{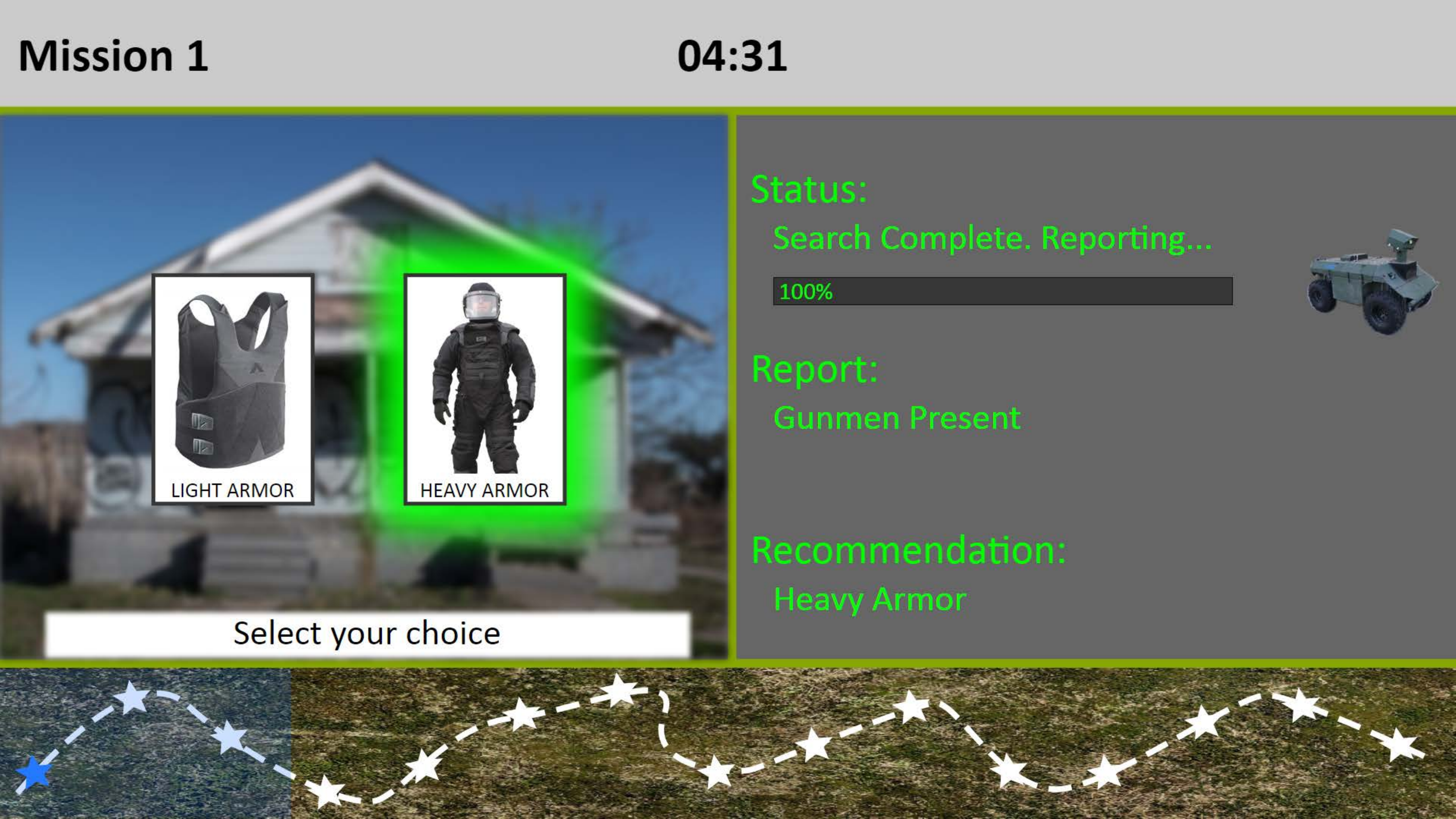}} \hfill
\subfigure[\label{fig_trans_mid}]{\includegraphics[width=0.32\textwidth]{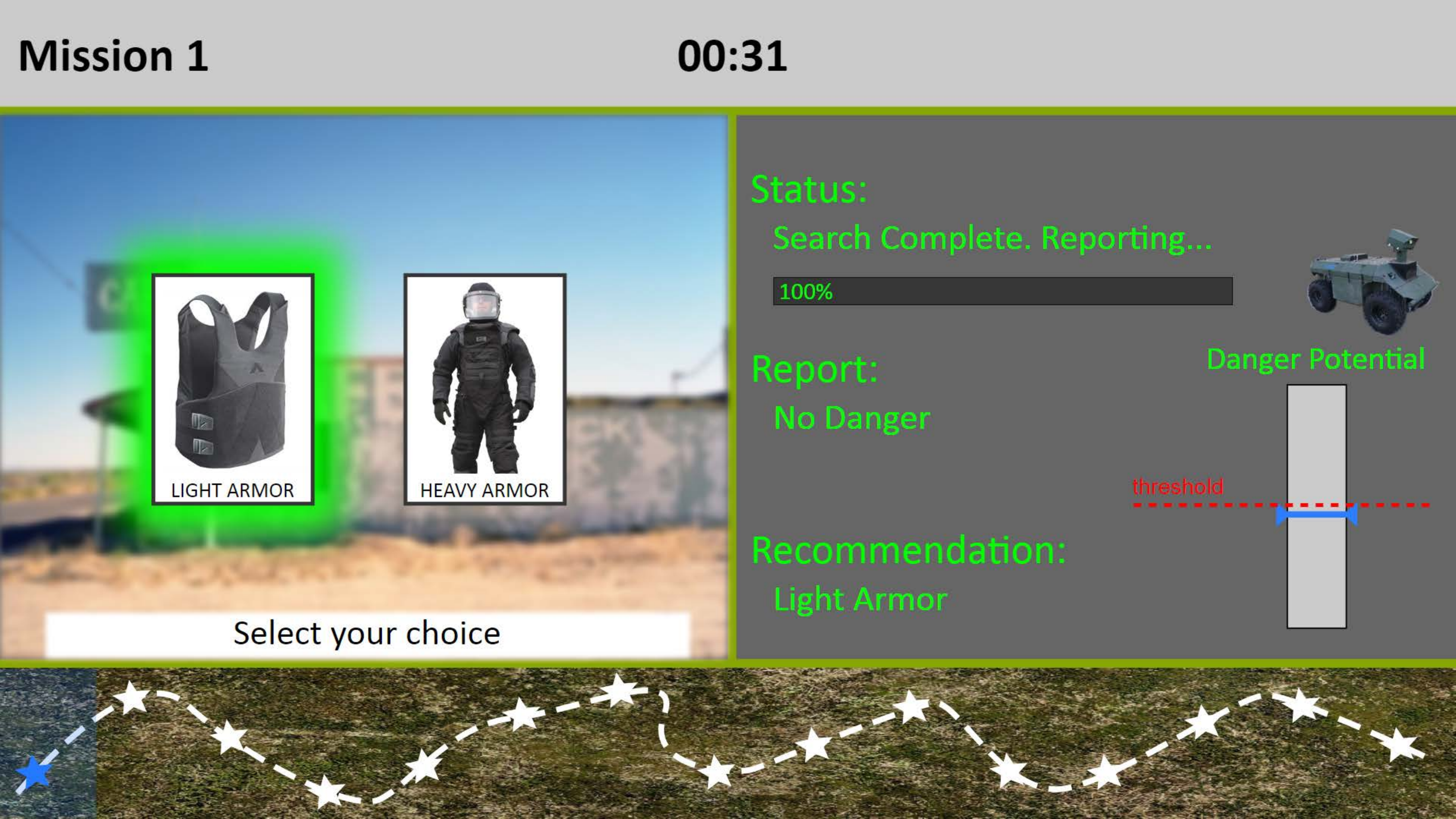}} \hfill
\subfigure[\label{fig_trans_high}]{\includegraphics[width=0.32\textwidth]{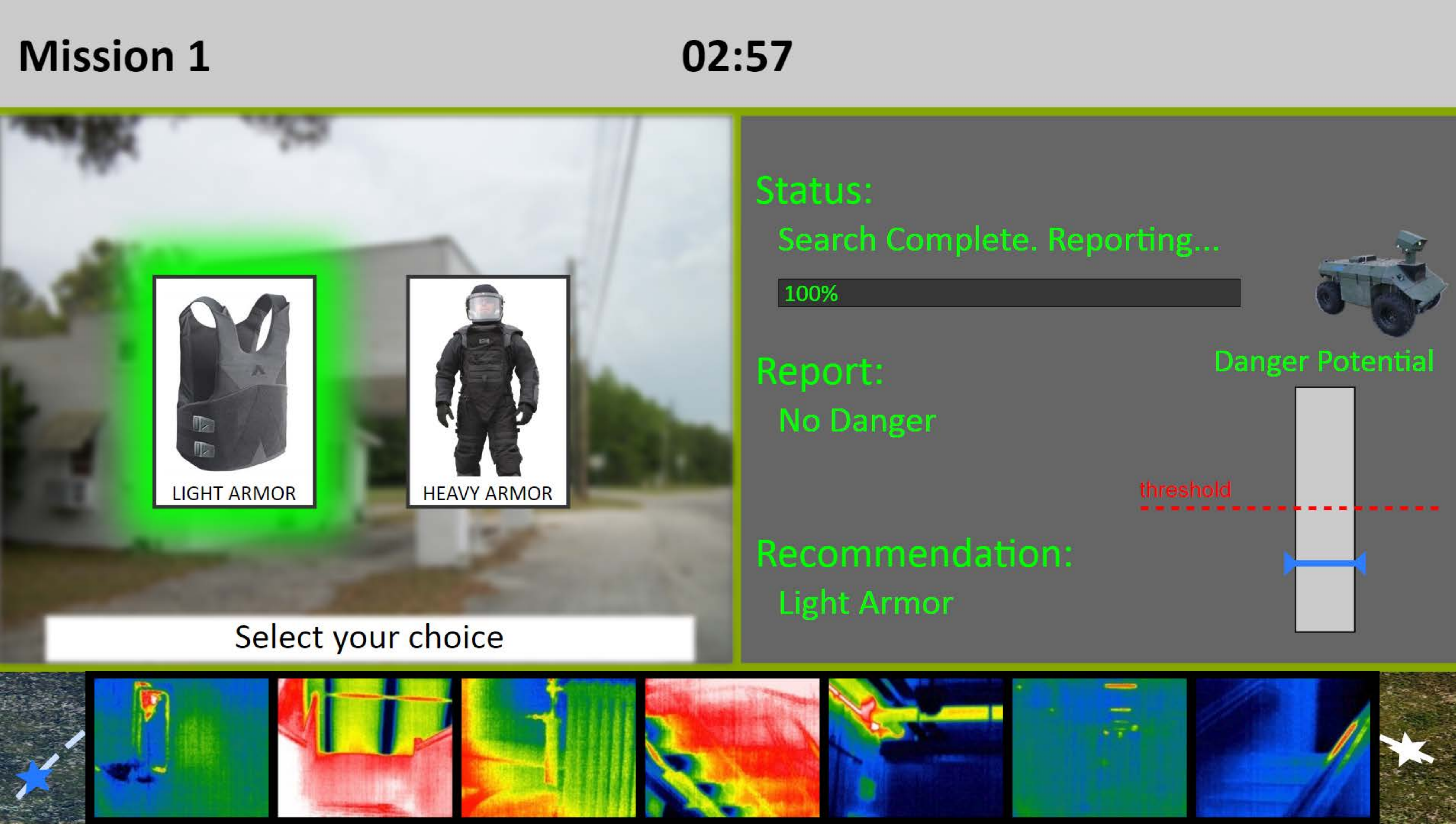}}
\caption{Example screenshots of robot reports corresponding to the three levels of transparencies. The top screenshot (a) shows a low transparency case with the robot's report (Gunmen Present) along with the armor recommendation (Heavy Armor). The middle screenshot (b) shows a medium transparency case that additionally includes a sensor bar on the left that indicates the level of potential danger perceived by the robot. The bottom screenshot (c) shows a high transparency case that further includes seven thermal images collected from inside the building, which the human can evaluate themselves.}
\label{fig_trans}
\end{figure}

In each mission, the participant was assisted by a robot that used a different transparency level for its recommendation. The interface for each of the levels of transparency is shown in Figure~\ref{fig_trans}. The low transparency robot reported if gunmen were present or absent along with the armor recommendation (see Figure~\ref{fig_trans_low}). The medium transparency robot additionally included a sensor bar indicating the level of potential danger as perceived by the robot (see Figure~\ref{fig_trans_mid}). The sensor reading was below the robot's threshold when no gunmen were detected and above the threshold when gunmen were detected. The high transparency robot included all of the information provided by the medium transparency robot, along with seven thermal images collected from inside the building (see Figure~\ref{fig_trans_high}). Note that this is only one way of defining different levels of transparency and can vary based on feasibility, context, and automation.

\begin{figure}
\centering
\includegraphics[width=0.58\textwidth]{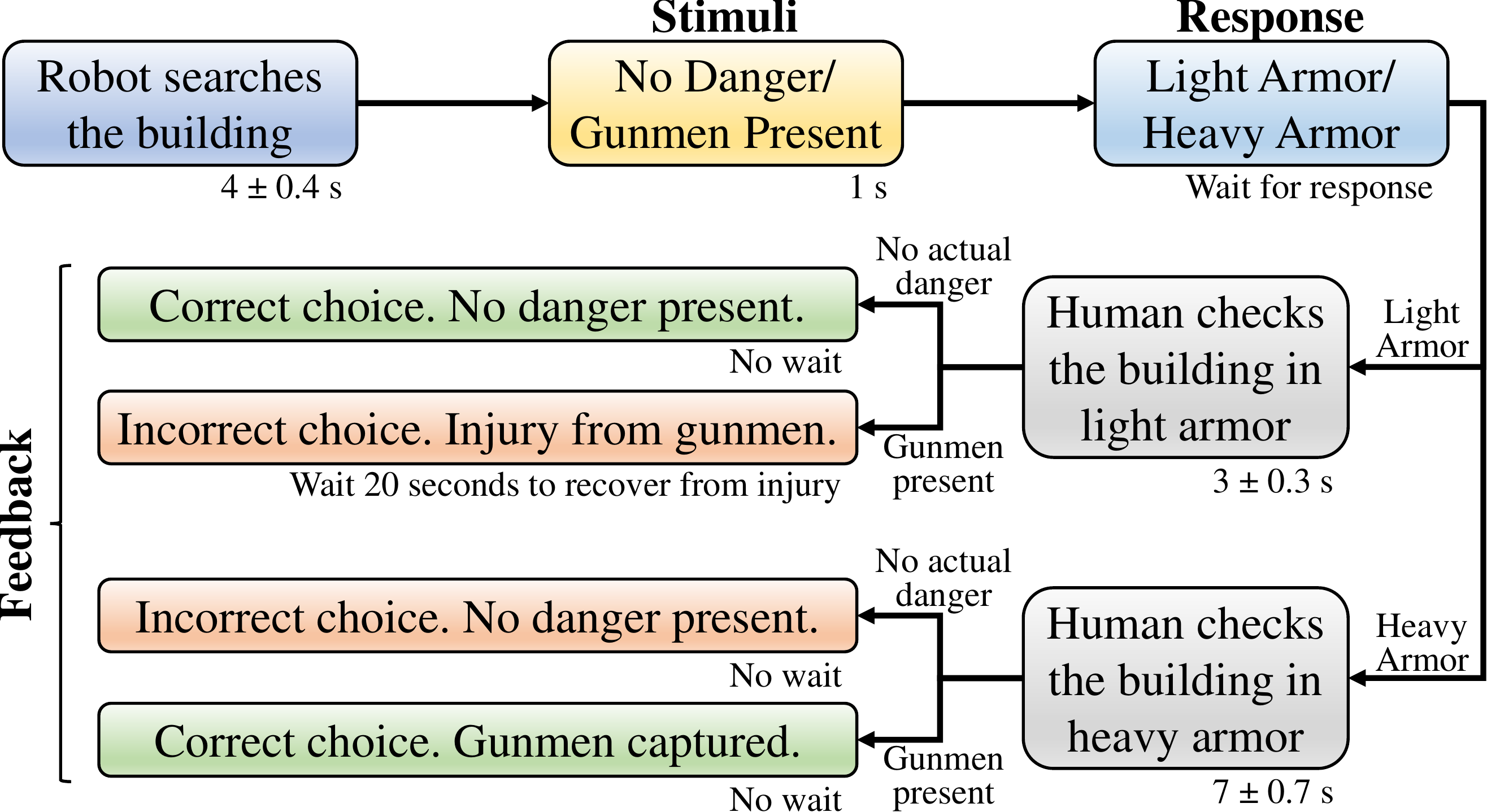}
\caption{\label{fig_Trial_Flow}The sequence of events in a single trial. The time length marked on the bottom right corner of each event indicates the time interval for which the information appeared on the computer screen.}
\end{figure}

Before the participants began the actual mission, they completed a tutorial mission consisting of six trials that helped familiarize them with the study interface and the three levels of transparency. The tutorial mission was uniform across all participants. For the experiment itself, the order of missions for each transparency level was randomized across participants to reduce ordering effects~\cite{shaughnessy2012research}. This randomization reduces the impact of factors like experience, practice from previous missions, and fatigue on the analysis. The presence or absence of gunmen was equally probable in each trial. The robots' recommendations were 70\% accurate. When the robot's recommendation was incorrect, it was a false alarm (false positive) or miss (false negative) with equal probability. The sequence of events in each trial is shown in Figure~\ref{fig_Trial_Flow}. 

\emph{Participants: }
Two hundred and twenty-one participants from the United States participated in, and completed, the study online. They were recruited using Amazon Mechanical Turk \cite{amazon2005amazon}, with the criteria that they must live in the US and have completed more than 1000 tasks with at least a 95\% approval rate. The compensation was \$1.50 for their participation, and each participant electronically provided their consent. The Institutional Review Board at Purdue University approved the study. Since the participants were not monitored while completing the study, we suspect that some participants were not sufficiently engaged with the study, reflected by their unusually high response times to stimuli. Therefore, we filtered data from participants who had any response time longer than the threshold at 99.5 percentile of all response times, which was approximately 40.45 seconds. As a result, 25 outlying participants were removed from the dataset.
\section{Model Parameter Estimation} \label{sec_estimate}

We assume that the trust and workload behavior of the general population can be represented by a common model. Therefore, we used the aggregated data from all participants to estimate the transition probability function, observation probability function, and the prior probabilities of states for the trust and workload models. For this study, the system recommendation that indicates \emph{Light Armor} is defined as Stimulus Absent~$S_A^-$ and the recommendation that indicates \emph{Heavy Armor} is defined as Stimulus Present~$S_A^+$.  We define a sequence of action-observation data for a participant as the interaction between the participant and robot in each mission. Therefore, we have $196\times 3$ sequences of data to estimate the parameters of each model.

The problem of model parameter estimation for POMDP models using sequences of data is defined as finding optimal parameters that maximize the likelihood of observing the sequences of observation for the given sequences of actions. For estimating the parameters of the discrete observation-space trust model, we use an extended version of the Baum-Welch algorithm, which is typically used for hidden Markov model (HMM) estimation (see \cite{rabiner1986introduction} for details). However, the continuous non-Gaussian distribution of the emission probability function of the workload model makes it infeasible to be estimated using the Baum-Welch algorithm. Therefore, we implement a genetic algorithm using Matlab to optimize the parameters for the workload model. The estimated POMDP models of trust and workload models are presented and analyzed in the next section.

\subsection{Trust Model}\label{subsec_trustmodel}

\begin{figure}
\centering
\includegraphics[width=\figsize]{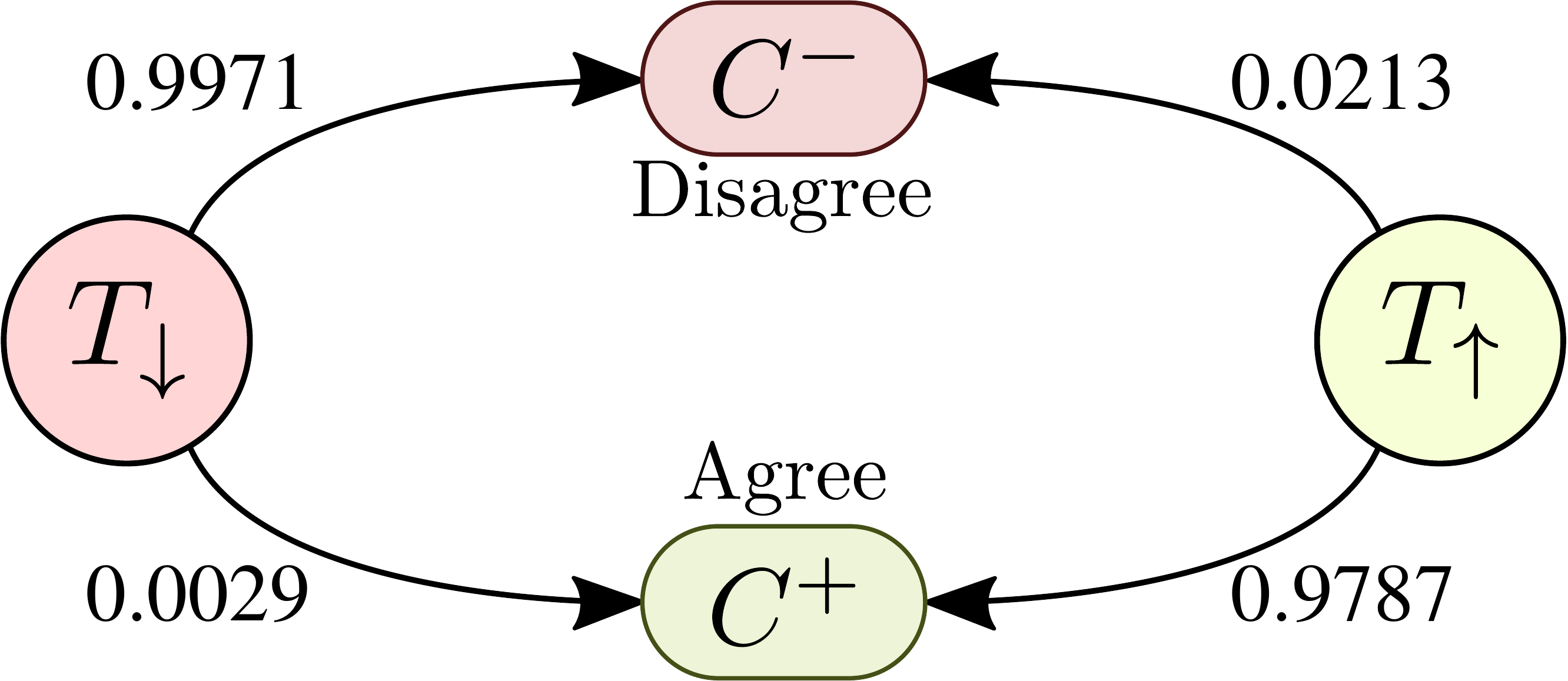}
\caption{Emission probability function $\mathcal{E}_T(o_C|s_T)$ for the trust model. Probabilities of observation are shown beside the arrows. Low Trust has a $99.71\%$ probability of resulting in participants disagreeing with the recommendation and High Trust has a $97.87\%$ probability of resulting in participants agreeing with the recommendation.}
\label{fig_trust_emission}
\end{figure}

The estimated trust model consists of initial state probabilities $\pi(s_T)$, an emission probability function $\mathcal{E}(o_C|s_T)$, and a transition probability function $\mathcal{T}(s_T'|s_T,a)$. The estimated initial probabilities of Low Trust $T_\downarrow$ and High Trust $T_\uparrow$ are $\pi(T_\downarrow) = 0.1286$ and $\pi(T_\uparrow) = 0.8714$, respectively.
This is consistent with findings that recent widespread use of automation has led to humans trusting a system when they have no experience with it \cite{merritt2008not}. The emission probability function $\mathcal{E}_T(o_C|s_T)$ is depicted in Figure~\ref{fig_trust_emission} and characterizes the probability of a participant's compliance with the system's recommendations given the participant's state of trust.
Both states, Low Trust and High Trust, have more than $97\%$ probability to result in participants disagreeing and agreeing with the recommendation, respectively. However, there is still a small probability of participants disagreeing while in a state of High Trust as well as participants agreeing while in a state of Low Trust. This inherently captures the uncertainty in human behavior.

\begin{figure}
\centering
\subfigure[\label{fig_Trust_trans_11}]{\includegraphics[width=0.35\textwidth]{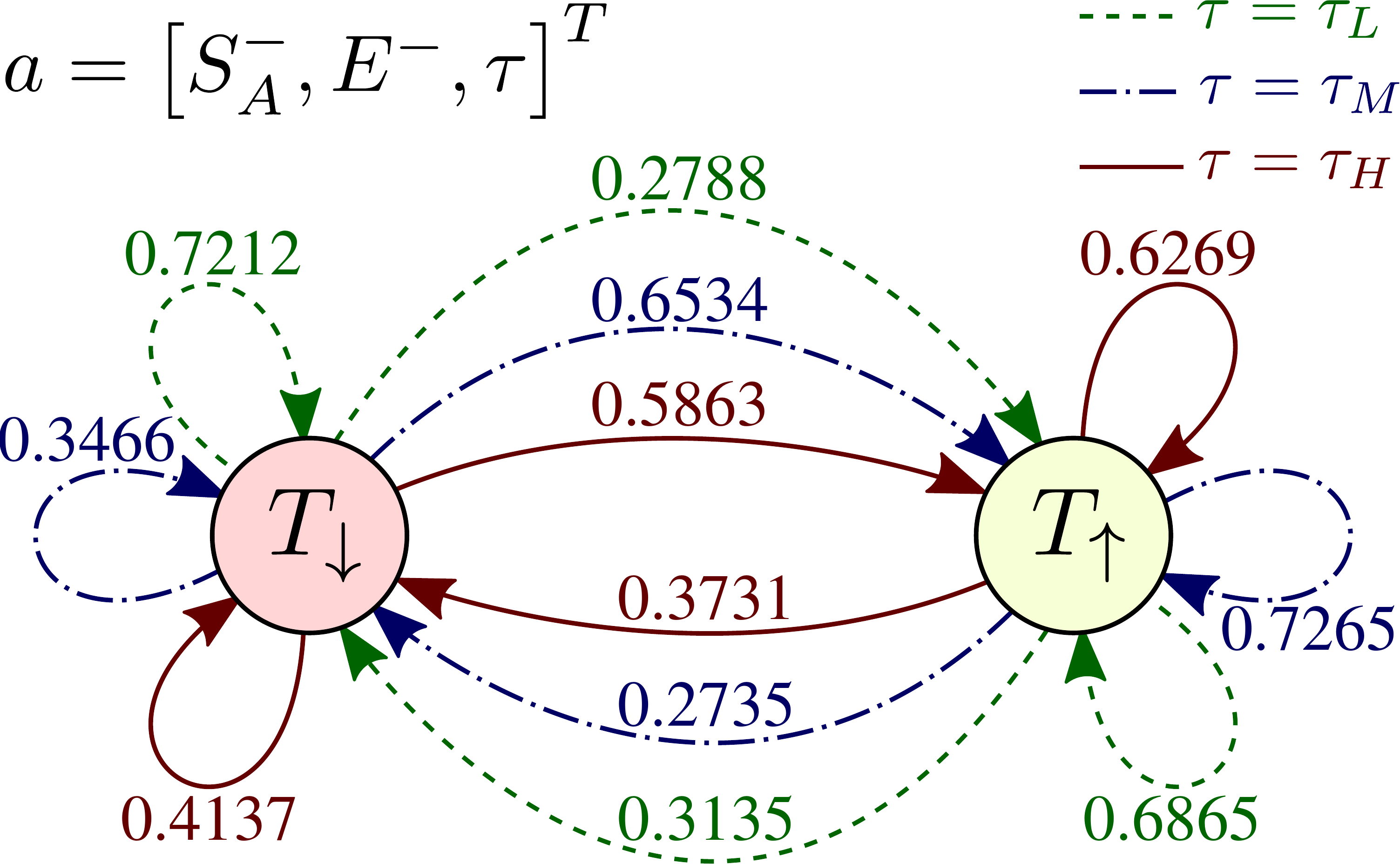}}
~
\subfigure[\label{fig_Trust_trans_12}]{\includegraphics[width=0.35\textwidth]{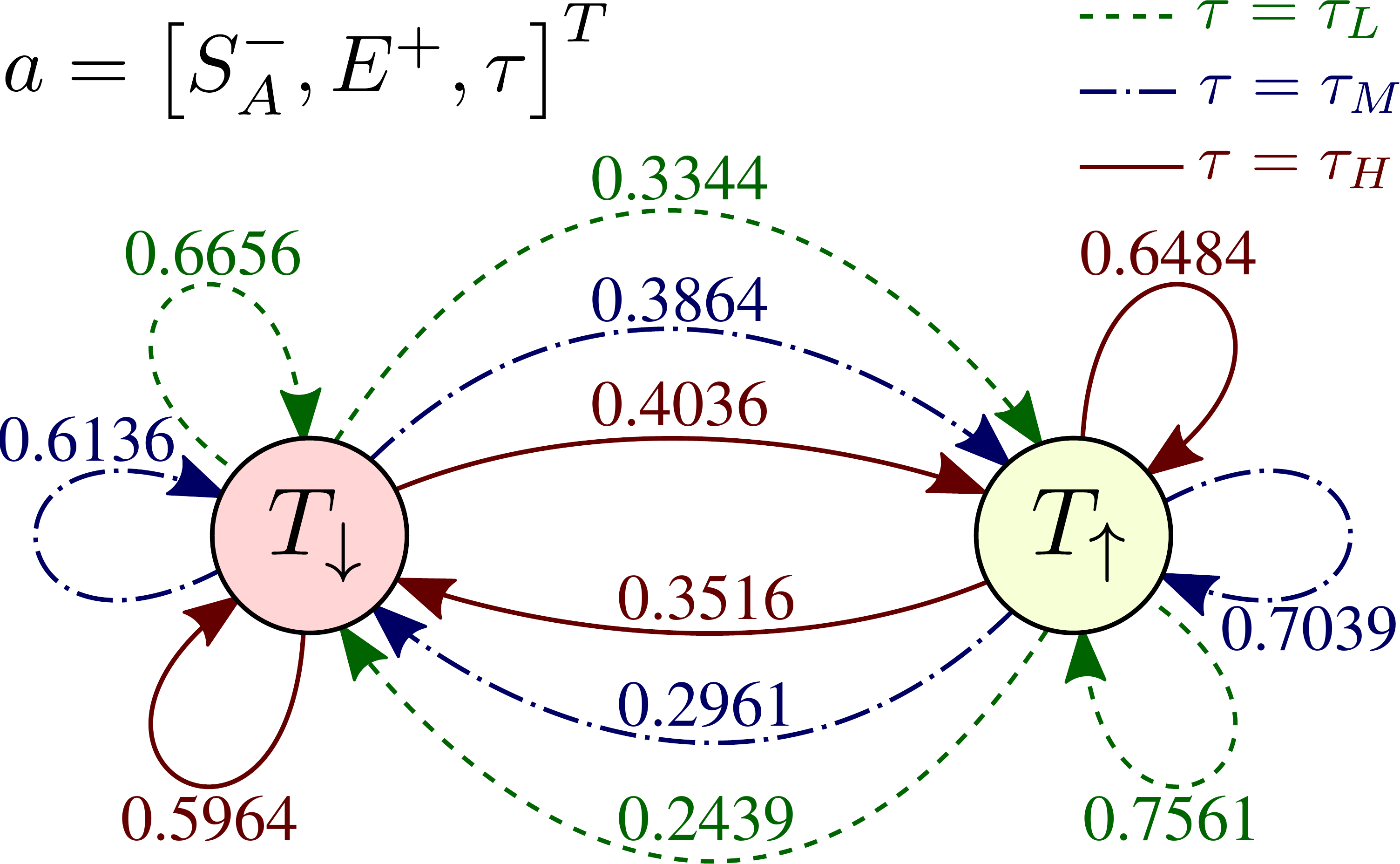}}
\\
\subfigure[\label{fig_Trust_trans_21}]{\includegraphics[width=0.35\textwidth]{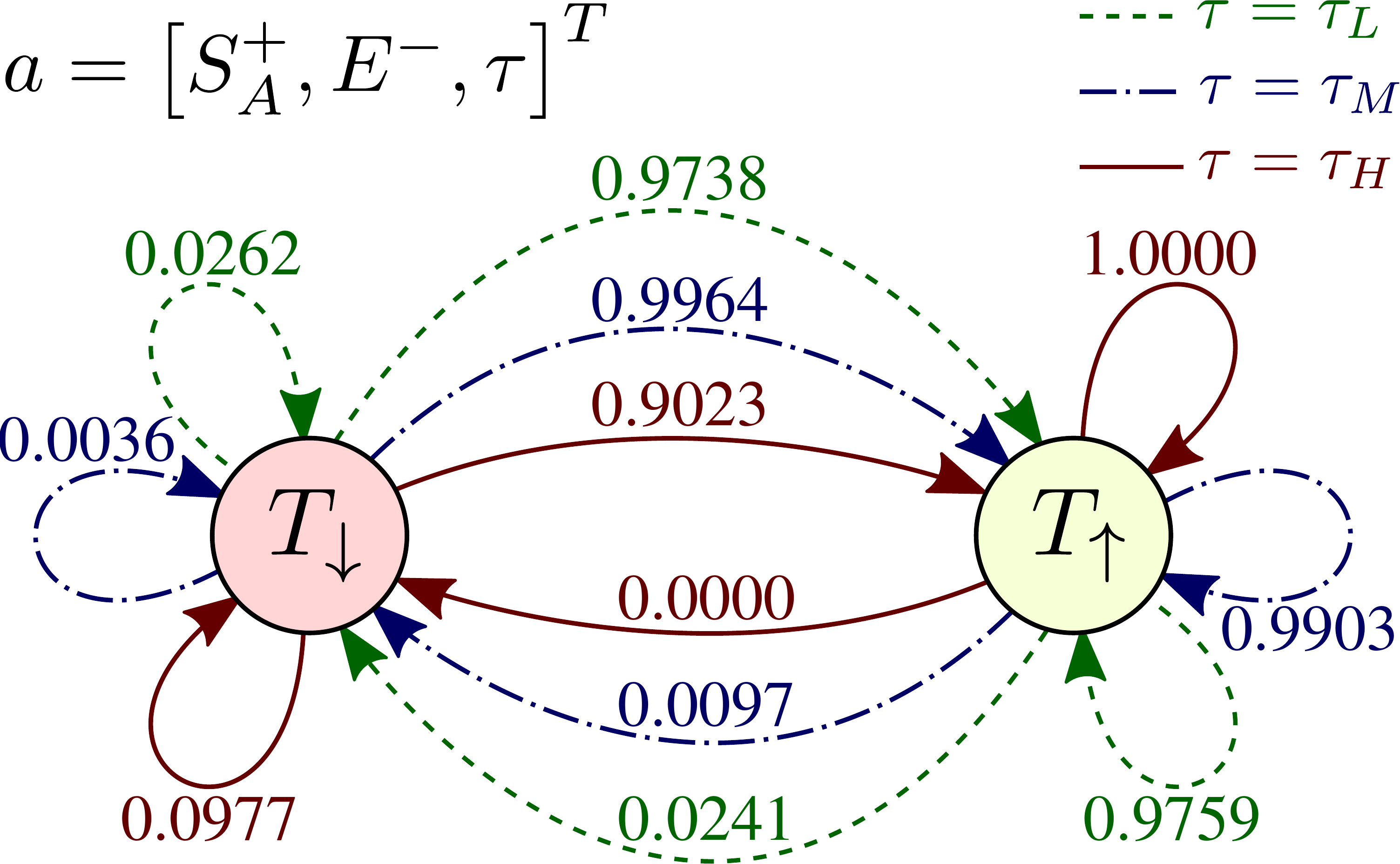}}
~
\subfigure[\label{fig_Trust_trans_22}]{\includegraphics[width=0.35\textwidth]{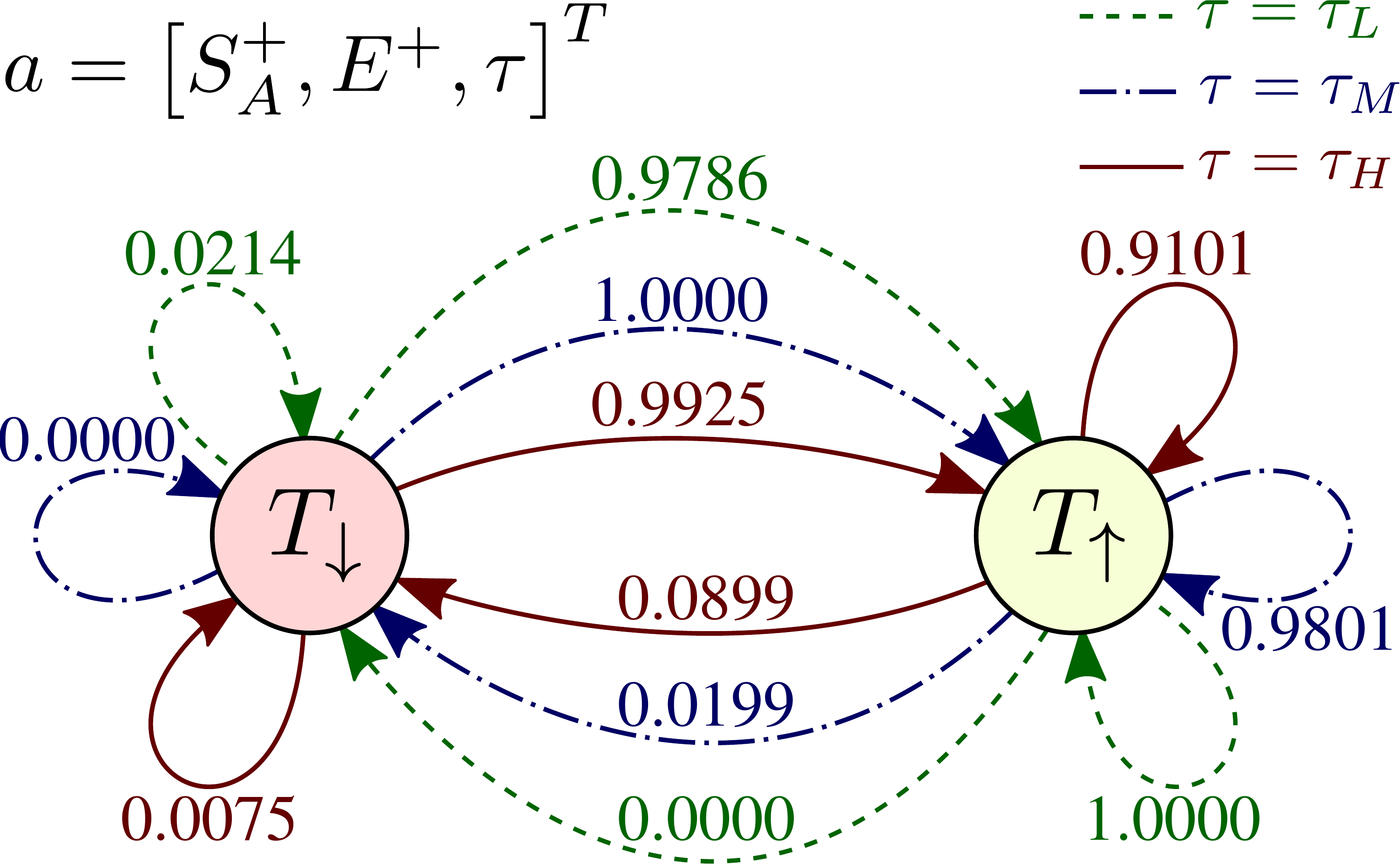}}
\caption{Transition probability function $\mathcal{T}_T(s_T'|s_T,a)$ for the trust model. Probabilities of transition are shown beside the arrows. The top-left diagram (a) shows the transition probabilities when the decision-aid's recommendation is Light Armor $S_A^-$ and the participant had a Faulty last experience $E^-$. The top-right diagram (b) shows the transition probabilities when the decision-aid recommends Light Armor $S_A^-$ and the participant had a Reliable last experience $E^+$. Both cases (a) and (b) can be considered relatively high-risk situations in this context because incorrectly complying with a faulty recommendation---that is, wearing Light Armor in the presence of gunmen---can result in injury. The bottom-left diagram (c) shows the transition probabilities when the decision-aid recommends Heavy Armor $S_A^+$ and the participant had a Faulty last experience $E^-$. The bottom-right diagram (d) shows the transition probabilities when the decision-aid recommends Heavy Armor $S_A^+$ and the participant had a Reliable last experience $E^+$.}
\label{fig_transition_trust}
\end{figure}

Figure~\ref{fig_transition_trust} represents the transition probability function $\mathcal{T}_T(s_T'|s_T,a)$ showing the probability of transitioning from the state $s_T$ to $s_T'$ (where $s_T,s_T' \in \set{T}$) given the action $a \in \mathcal{A}$.
The cases when the recommendation suggests Light Armor $S_A^-$ can be considered relatively high-risk situations in our context because incorrectly complying with a faulty recommendation---that is, wearing Light Armor in the presence of gunmen---can result in getting injured and a penalty of 20 seconds (see (Figures~\ref{fig_Trust_trans_11} and \ref{fig_Trust_trans_12}). On the other hand, the cases when the recommendation suggests Heavy Armor $S_A^+$ are low-risk situations (Figures~\ref{fig_Trust_trans_21} and \ref{fig_Trust_trans_22}) as incorrect compliance only leads to an extra 4 seconds of search time.  We observe that the probability of transitioning to High Trust $T_\uparrow$ as well as staying in  High Trust $T_\uparrow$ is higher for low-risk situations (Figure~\ref{fig_Trust_trans_21} and \ref{fig_Trust_trans_22}) as compared to the corresponding high-risk situations (Figure~\ref{fig_Trust_trans_11} and \ref{fig_Trust_trans_12}). Given that the robot was only $70\%$ reliable in the study, this over-trust during low-risk situations, with transition probabilities to High Trust $T_\uparrow$ being greater than $91\%$ (see Figure~\ref{fig_Trust_trans_21} and \ref{fig_Trust_trans_22}), indicates the inherent conservative behavior of participants. The participants preferred to comply with the robot (by choosing Heavy Armor) and risk an effective penalty of 4 seconds instead of risking a penalty of 20 seconds. It should be noted that the participants did not know about the failure rate of the robot.

Interestingly, we observe that high transparency $\tau_H$ has the highest probability of causing a transition from High Trust $T_\uparrow$ to Low Trust $T_\downarrow$ as compared to lower transparencies in most cases (except Figure~\ref{fig_Trust_trans_21}). This is because high transparency enables the participant to make a more informed decision and avoid errors that would result from trusting a faulty recommendation. Therefore, high transparency helps the participant to calibrate their trust correctly in these cases. Moreover, in most cases (except Figure~\ref{fig_Trust_trans_21}), low transparency $\tau_L$ has the lowest probability of causing a transition from Low Trust $T_\downarrow$ to High Trust $T_\uparrow$. However, low transparency can offer the best strategy for maintaining a state of high trust (see Figure~\ref{fig_Trust_trans_12}). In summary, our findings suggest that transparency does not directly affect human trust; instead, factors such as the human's current trust state, assessment of risk, and the reliability of the automation affect how transparency changes trust.

\subsection{Workload Model}
\begin{figure}
\centering
\includegraphics[width=.50\textwidth]{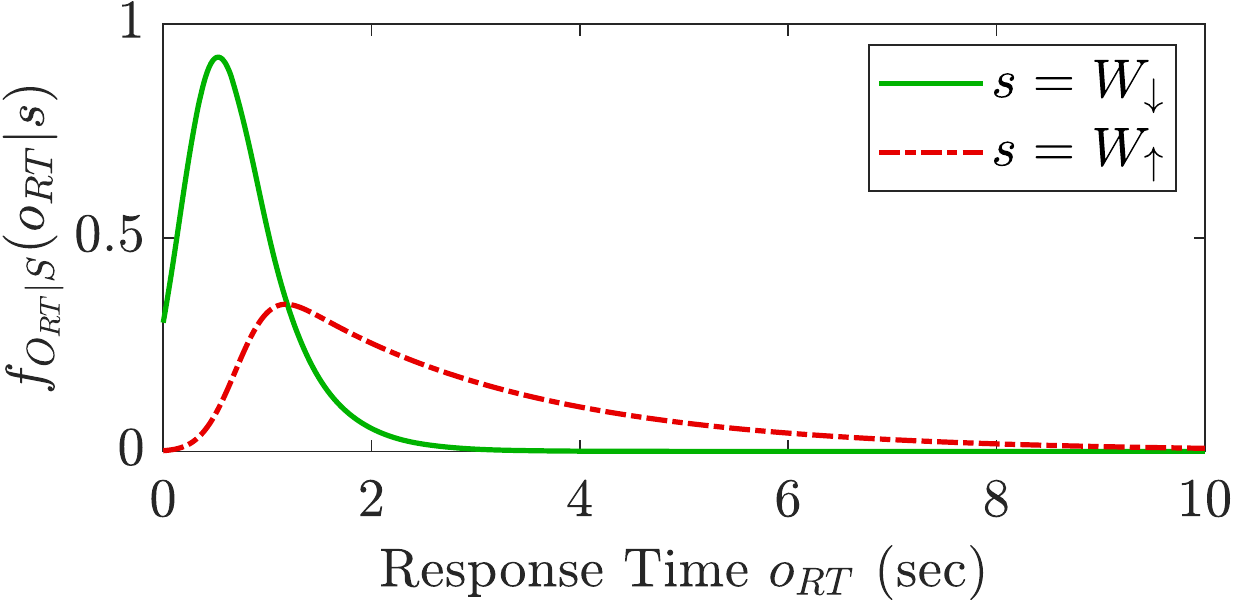}
\caption{Emission probability function $\mathcal{E}_W(o_{RT}|s_W)$ for the workload model. For Low Workload, the response time ($o_{RT}$) probability density function $f_{O_{RT}|W_\downarrow}(o_{RT}|W_\downarrow)$ is characterized by an ex-Gaussian distribution with $\mu_{W_\downarrow} = 0.2701$, $\sigma_{W_\downarrow} = 0.2964$, and $\tau_{W_\downarrow} = 0.4325$. For High Workload, the response time ($o_{RT}$) probability density function $f_{O_{RT}|W_\uparrow}(o_{RT}|W_\uparrow)$ is characterized by an ex-Gaussian distribution with $\mu_{W_\uparrow} = 0.7184$, $\sigma_{W_\uparrow} = 0.2689$, and $\tau_{W_\uparrow} = 2.2502$. Low Workload $W_\downarrow$ is more likely than High Workload to result in a response time of less than approximately 1.19 seconds.}
\label{fig_emiss_Workload}
\end{figure}

We estimated the initial probabilities of Low Workload $W_\downarrow$ and High Workload $W_\uparrow$ to be $\pi(W_\downarrow) = 0.3097$ and $\pi(W_\uparrow) = 0.6903$, respectively.
The high initial probability of High Workload $W_\uparrow$ is expected because participants initially need to familiarize themselves with the system.
The emission probability function $\mathcal{E}_W(o_{RT}|s_W)$ is represented in Figure~\ref{fig_emiss_Workload}, which shows the probability density function of observing participants' response time as $o_{RT}$ given their state of workload $s_W$.
We observe that Low Workload $W_\downarrow$ is more likely than High Workload to result in a response time of less than approximately 1.19 seconds. High Workload is more likely to lead to high response times.

\begin{figure}
\centering
\subfigure[\label{fig_Workload_trans_11}]{\includegraphics[width=0.35\textwidth]{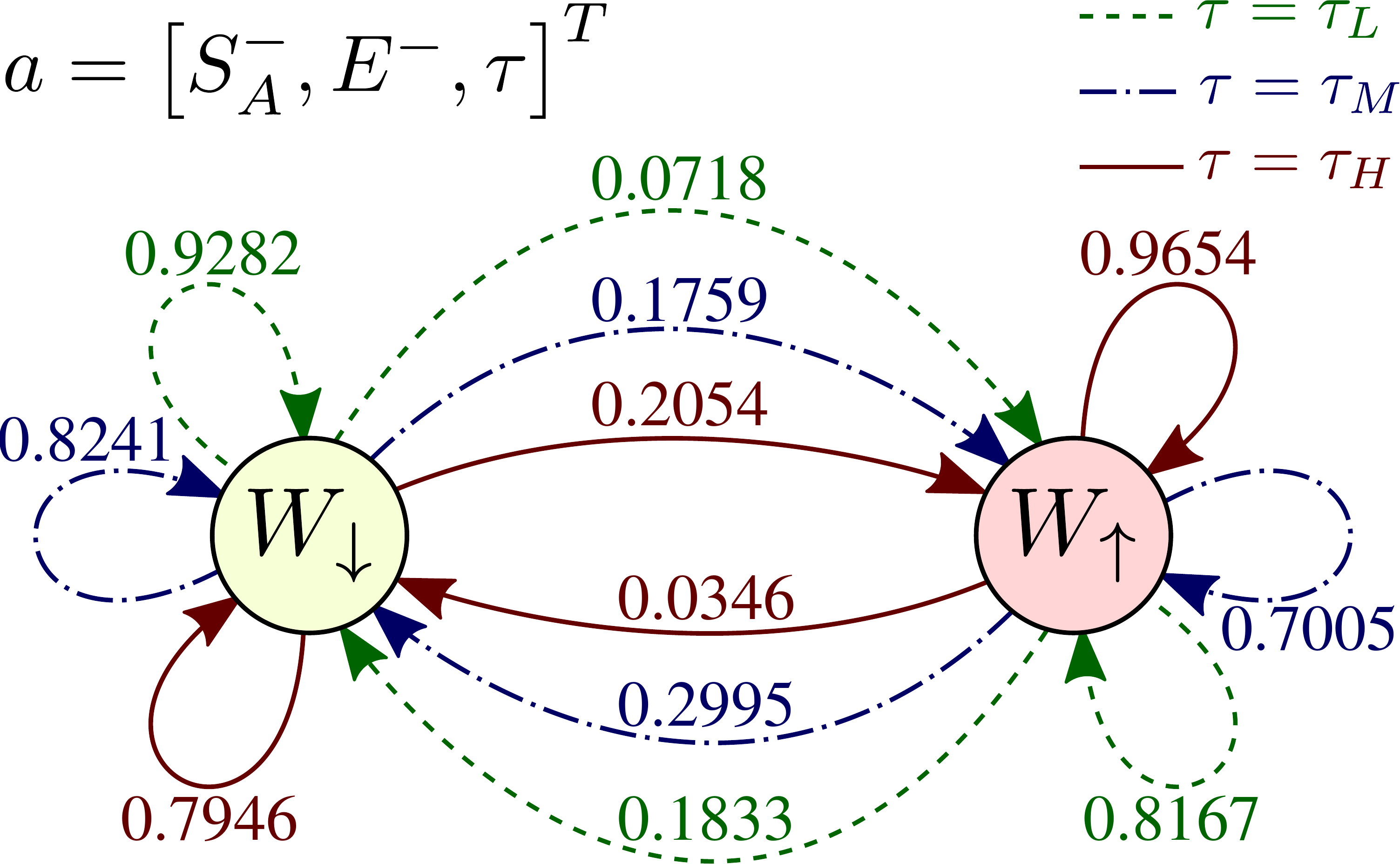}}
~
\subfigure[\label{fig_Workload_trans_12}]{\includegraphics[width=0.35\textwidth]{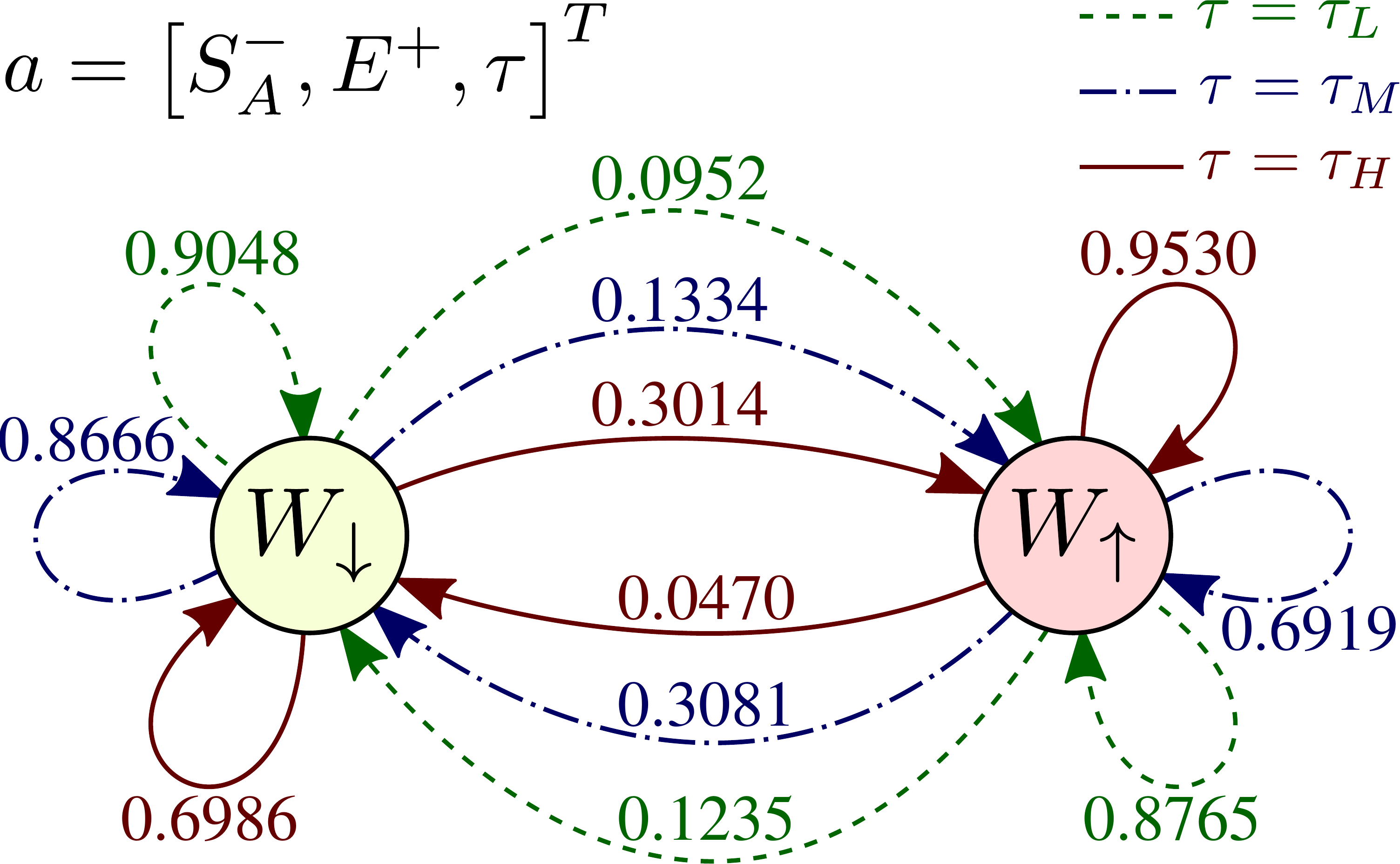}}
\\
\subfigure[\label{fig_Workload_trans_21}]{\includegraphics[width=0.35\textwidth]{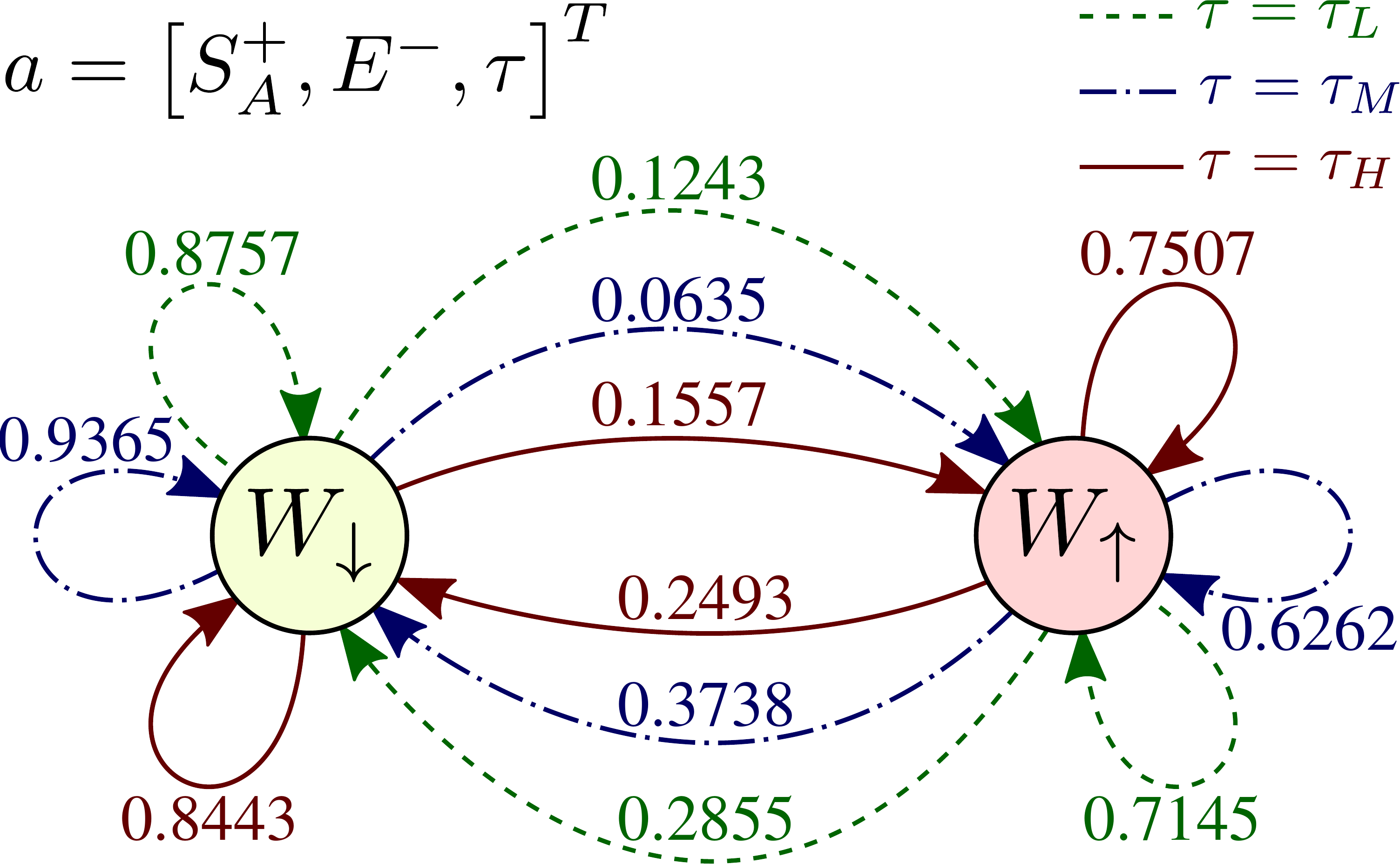}}
~
\subfigure[\label{fig_Workload_trans_22}]{\includegraphics[width=0.35\textwidth]{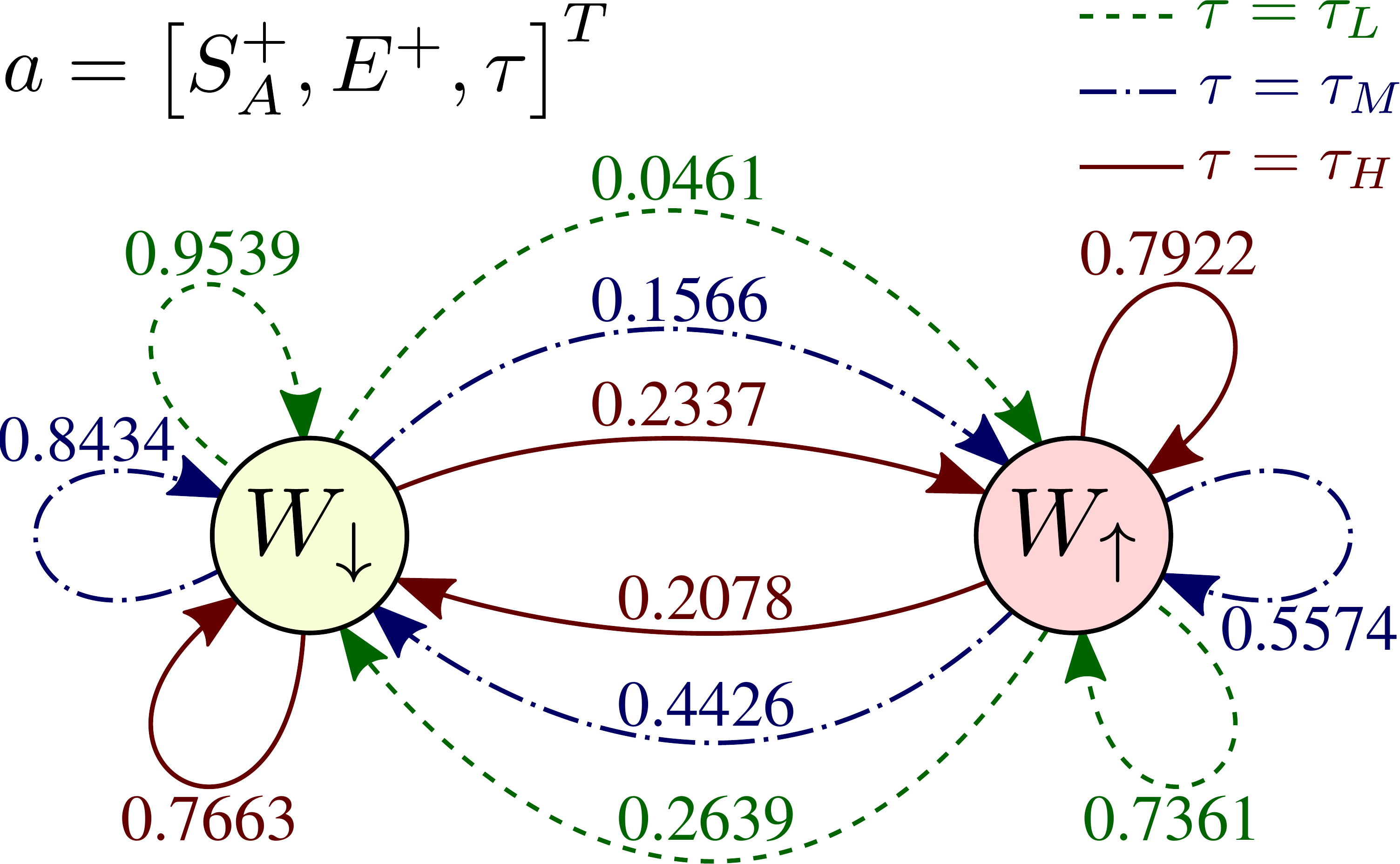}}
\caption{Transition probability function $\mathcal{T}_W(s_W'|s_W,a)$ for the workload model. Probabilities of transition are shown beside the arrows. The top-left diagram (a) shows the transition probabilities when the decision-aid recommends Light Armor $S_A^-$ and the participant had a Faulty last experience $E^-$. The top-right diagram (b) shows the transition probabilities when the decision-aid recommends Light Armor $S_A^-$ and the participant had a Reliable last experience $E^+$. The bottom-left diagram (c) shows the transition probabilities when the decision-aid recommends Heavy Armor $S_A^+$ and the participant had a Faulty last experience $E^-$. The bottom-right diagram (d) shows the transition probabilities when the decision-aid recommends Heavy Armor $S_A^+$ and the participant had a Reliable last experience $E^+$.}
\label{fig_transition_Workload}
\end{figure}

The transition probability function $\mathcal{T}_W(s_W'|s_W,a)$ is represented in Figure~\ref{fig_transition_Workload} and shows the probability of a participant transitioning from the state $s_W$ to $s_W'$ based on the action $a \in \mathcal{A}$, where $s_W,s_W' \in \set{W}$.
We observe that in most cases, the probability of transitioning from Low Workload $W_\downarrow$ to High Workload $W_\uparrow$ is greater for higher transparencies for a given recommendation and experience (except Figure~\ref{fig_Workload_trans_21}). Therefore, it is more likely that higher transparencies will increase participants' workload if they are in a state of Low Workload $W_\downarrow$ because they need to process even more information for decision-making. However, if a participant is in a state of High Workload $W_\uparrow$, medium transparency $\tau_M$ has a lower probability than low transparency $\tau_L$ to keep them in a High Workload $W_\uparrow$ state. This may occur because in such cases, low transparency may not provide enough information for the participants to make a decision, leaving them confused. This results in participants using more time and effort to reach a decision. Overall, high transparency $\tau_H$ has the highest probability of keeping the participant in a state of High Workload $W_\uparrow$ for a given recommendation and experience. Finally, it is worth noting that the probability of transitioning to High Workload $W_\uparrow$ from any workload state is higher when the decision-aid recommends Light Armor $S_A^-$ (see Figure~\ref{fig_Workload_trans_11} and \ref{fig_Workload_trans_12}) as compared to when the decision-aid recommends Heavy Armor $S_A^+$ (see Figure~\ref{fig_Workload_trans_21} and \ref{fig_Workload_trans_22}) for a given experience and transparency. This is because a recommendation suggesting Light Armor $S_A^-$ has a higher risk as discussed previously, leading  humans to consider their decision more carefully.

In summary, we have created a POMDP model for estimating human trust and workload in the context of a human interacting with a decision-aid system. We observe that a higher transparency is not always the most likely way to increase trust in humans nor is it always more likely to increase workload. Instead, the optimal transparency depends on the current state of human trust and workload along with the recommendation type and the human's past experiences. In other words, higher transparency is not always beneficial, and instead, system transparency should be controlled based upon all these factors. In the next section, we use the POMDP model to develop an optimal control policy that varies system transparency to improve human-machine interaction performance objectives. 
\section{Controller Design} \label{sec_control}
 
Before we synthesize an optimal control policy, we need to define the context-specific performance objectives relevant in this study. We focus on two critical performance objectives: 1) the human should make correct decisions irrespective of the robot's reliability and 2) the human should make their decision in the shortest amount of time. Based on these performance objectives, we define the reward function for the POMDP, which is used to obtain the optimal control policy. 

\subsection{Decision Reward Function} 
The primary goal of calibrating trust and workload during human-machine interactions is to achieve the goals that are specific to the interaction. Since any given decision-aid system is never completely reliable, it is not always beneficial for the human to comply with the system. Instead, the human should make \emph{correct} decisions; that is, the human should comply with the system when its recommendation is reliable and not comply when the system's recommendation is faulty. The decision reward function aims to enforce this behavior by appropriately penalizing the human's decisions. In order to characterize this behavior formally, we first define a few terms. 

Earlier we defined the recommendation $a_{S_A} \in \set{S}_A \defeq \{S_A^-, S_A^+ \}$ of the decision-aid system depending on its inference about a given situation. However, we also need to distinguish the human's inference about the situation from the true situation. For example, in our reconnaissance mission, it is possible for the decision-aid robot to correctly recommend the use of Light Armor, indicating the absence of gunmen, i.e., $S_A^-$, but for the human to believe that the robot is faulty.  In this case, the human may infer that there are gunmen present and choose to wear Heavy Armor. To account for this situation, we additionally define the true absence or presence of the stimulus as $\bar{a}_S \in \set{S} \defeq \{S^-,S^+\}$ and the human's inference as $\bar{a}_{S_H} \in \set{S}_H \defeq \{S_H^-,S_H^+\}$. Note that terms with $\bar{a}$ should not be confused with actions $a$ of the POMDP model. Also, $\sbt^-$ and $\sbt^+$ represent the absence and presence of a stimulus, respectively. Typically, the prior probability of the true situation $\pr(\bar{a}_S)$ is known. Moreover, the presence or absence of gunmen is equally probable in our study, so $\pr(S^-)=\pr(S^+)=0.5$.

The decision-aid's recommendations, and the human's decisions with respect to the true situation, are each characterized by the confusion matrix shown in Table~\ref{tab_confusion}.
\begin{table}
    \caption{Confusion matrix representation for the decision-aid system's and the human's inference. Each row of the matrix represents the true situation, while each column represents the inference made by the decision-aid system or the human.} \label{tab_confusion}
    \centering
    \begin{tabular}{cc|c|c|}
        \cline{3-4}
        \multicolumn{2}{c|}{\multirow{2}{*}{}}                         & \multicolumn{2}{c|}{Decision-aid system's or human's inference}                     \\ \cline{3-4} 
        \multicolumn{2}{c|}{}                                          & $a_{S_A} = S_A^-$ or $\bar{a}_{S_H} = S_H^-$                                          & \begin{tabular}[c]{@{}c@{}}$a_{S_A} = S_A^+$ or $\bar{a}_{S_H} = S_H^+$\end{tabular} \\ \hline
        \multicolumn{1}{|c|}{\multirow{2}{*}{True Situation}} & $\bar{a}_S = S^-$ & \begin{tabular}[c]{@{}c@{}}True Negative\\ TN\end{tabular}  & \begin{tabular}[c]{@{}c@{}}False Positive\\ FP\end{tabular}   \\ \cline{2-4} 
        \multicolumn{1}{|c|}{}                                & $\bar{a}_S = S^+$ & \begin{tabular}[c]{@{}c@{}}False Negative\\ FN\end{tabular} & \begin{tabular}[c]{@{}c@{}}True Positive\\ TP\end{tabular}    \\ \hline
    \end{tabular}
\end{table}
In practice, a decision-aid system's reliability is a system characteristic and known \emph{a priori}; therefore, we define the reliability function as a probability of the system's recommendation given the true situation, i.e., $p(a_{S_A}|\bar{a}_S)$; we denote the probability of the decision-aid system making a false negative as $\pr(S_A^-|S^+)=\beta$  and  the probability of the decision-aid system making a false positive as $\pr(S_A^+|S^-)=\alpha$. These reliability characteristics of the decision-aid system with $70\%$ reliability in our reconnaissance mission study are summarized in Table~\ref{tab_machine_reliability}.
\begin{table}
    \caption{Reliability characteristics of the decision-aid system in the reconnaissance mission study representing the probabilities of the decision-aid's inference given the true situation. Since the decision-aid is $70\%$ reliable, the probability of the decision-aid making a correct inference is $0.7$.} \label{tab_machine_reliability}
    \centering
    \begin{tabular}{cc|c|c|}
        \cline{3-4}
        \multicolumn{2}{c|}{\multirow{2}{*}{}}                         & \multicolumn{2}{c|}{Decision-aid robot's inference}                     \\ \cline{3-4} 
        \multicolumn{2}{c|}{}                                          & $a_{S_A} = S_A^-$                                           & \begin{tabular}[c]{@{}c@{}}$a_{S_A} = S_A^+$ \end{tabular} \\ \hline
        \multicolumn{1}{|c|}{\multirow{2}{*}{True Situation}} & $\bar{a}_S = S^-$ & \begin{tabular}[c]{@{}c@{}}$1-\alpha$\\ $=0.7$\end{tabular}  & \begin{tabular}[c]{@{}c@{}}$\alpha$\\ $=0.3$\end{tabular}   \\ \cline{2-4} 
        \multicolumn{1}{|c|}{}                                & $\bar{a}_S = S^+$ & \begin{tabular}[c]{@{}c@{}}$\beta$\\ $=0.3$\end{tabular} & \begin{tabular}[c]{@{}c@{}}$1-\beta$\\ $=0.7$\end{tabular}    \\ \hline
    \end{tabular}
\end{table}
To help the human make correct decisions, we define a decision reward function $\mathcal{R}_{D}: \set{S}_H \times \set{S} \rightarrow \mathbb{R}$ in terms of the human inference and the true situation, which is summarized in Table~\ref{tab_rewards}.
\begin{table}
    \caption{Decision reward function based on the inference made by the human. The reward function is defined as penalties equivalent to the expected amount of time, in seconds, that the human has to expend as a result of their decision.} \label{tab_rewards}
    \centering
    \begin{tabular}{cc|c|c|}
        \cline{3-4}
        \multicolumn{2}{c|}{\multirow{2}{*}{}}                         & \multicolumn{2}{c|}{Human's inference}                     \\ \cline{3-4} 
        \multicolumn{2}{c|}{}                                          & $\bar{a}_{S_H} = S_H^-$                                           & \begin{tabular}[c]{@{}c@{}}$\bar{a}_{S_H} = S_H^+$ \end{tabular} \\ \hline
        \multicolumn{1}{|c|}{\multirow{2}{*}{True Situation}} & $\bar{a}_S = S^-$ & \begin{tabular}[c]{@{}c@{}}$-3$\end{tabular}  & \begin{tabular}[c]{@{}c@{}}$-7$\end{tabular}   \\ \cline{2-4} 
        \multicolumn{1}{|c|}{}                                & $\bar{a}_S = S^+$ & \begin{tabular}[c]{@{}c@{}}$-23$\end{tabular} & \begin{tabular}[c]{@{}c@{}}$-7$\end{tabular}    \\ \hline
    \end{tabular}
\end{table}
The reward function is defined in terms of penalties equivalent to the expected amount of time, in seconds, that the human has to expend as a result of their decision. In particular, the human has to wait 3 seconds to search the building with Light Armor $S_H^-$ if there are no gunmen present $S^-$. However, if gunmen are present $S^+$ and the human chooses Light Armor $S_H^-$, an additional 20 second penalty due to injury will be applied, resulting in a total wait time of 23 seconds. Moreover, a choice of Heavy Armor $S_H^+$ will always result in a wait of 7 seconds to search the building irrespective of the true situation. This reward function is specific to the reconnaissance study context and should be, in general, defined based on the context under consideration.

Although, the decision reward function $\mathcal{R}_{D}(\bar{a}_{S_H},\bar{a}_S)$ is intuitive to design, the standard form of the reward function for a POMDP $\mathcal{R}: \mathcal{S} \times \mathcal{S} \times \mathcal{A} \rightarrow \mathbb{R}$ is defined as the reward for transitioning from state $s\in\mathcal{S}$ to $s'\in\mathcal{S}$ due to action $a\in\mathcal{A}$. Therefore, we transform $\mathcal{R}_{D}(\bar{a}_{S_H},\bar{a}_S)$ to derive the expected standard reward function. As decision rewards are only dependent on human compliance behavior and therefore only on trust behavior, we derive the expected reward function for the trust POMDP model as $\mathcal{R}_T: \set{T} \times \set{T} \times \mathcal{A} \rightarrow \mathbb{R}$ by calculating $\E{R|s_T,s_T',a}$, where random variable $R$ is the reward and $\E{\sbt}$ is the expected value of $\sbt$. 

Given the reward function $\mathcal{R}_{D}(\bar{a}_{S_H},\bar{a}_S)$ as shown in Table~\ref{tab_rewards}, trust emission probability function $\mathcal{E}_T(o_C,s_T)$, and automation reliability characteristics defined as $p(a_{S_A}|\bar{a}_S)$, an equivalent expected reward function in the form $\mathcal{R}_T(s_T,s_T',a)$ is 
\begin{align} \label{eq_trust_rewards}
    \mathcal{R}_T(s_T,s_T',[a_{S_A},a_E,a_\tau]) = \sum\limits_{o_C'\in C} \sum\limits_{\bar{a}_S\in S} \mathcal{E}_T\left(o_C'|s_T'\right) \pr\left(\bar{a}_S|a_{S_A}\right)\mathcal{R}_D\left(g\left(a_{S_A},o_C'\right),\bar{a}_S\right) \enspace ,
\end{align}
where $\pr\left(\bar{a}_S|a_{S_A}\right)$ is the posterior probability calculated from Table~\ref{tab_machine_reliability} and Bayes' theorem as $$\pr\left(\bar{a}_S|a_{S_A}\right) = \frac{\pr\left(a_{S_A}|\bar{a}_S\right) \pr\left(\bar{a}_S\right)}{\sum\limits_{\bar{a}_S\in S} \pr\left(a_{S_A}|\bar{a}_S\right) \pr\left(\bar{a}_S\right)} \enspace,$$ and $g:S_A\times C \rightarrow S_H$ is a function mapping human compliance $o_C \in C$ in response to the system's recommendation $a_{S_A}\in S_A$ to the human inference/decision $\bar{a}_{S_H}\in S_H$. For example, the human not complying $o_C = C^-$ with a recommendation of Light Armor $a_{S_A}= S_A^-$ effectively means that the human is inferring the presence of gunmen, and thereby choosing Heavy Armor $\bar{a}_{S_H}= S_H^+$. Specifically,
$$g\left(S_A^-,C^- \right) = S_H^+ \enspace,\enspace g\left(S_A^-,C^+ \right) = S_H^- \enspace,\enspace g\left(S_A^+,C^- \right) = S_H^- \enspace,\enspace \text{and} \enspace g\left(S_A^+,C^+ \right) = S_H^+ \enspace. $$

\subsection{Response Time Reward Function} 
In most scenarios, apart from ensuring that the human makes correct decisions, the time the human takes to make the decision is also critical.
Therefore, to minimize the human's response time, we define the response time reward function  $\mathcal{R}_{RT}:  \mathbb{R}^+ \rightarrow \mathbb{R}$ as $\mathcal{R}_{RT} \left(o_{RT} \right) = -o_{RT}$ to proportionally penalize longer response times. Similar to the decision reward function, the response time reward function is transformed to derive the expected standard reward function for the POMDP. As the response time reward function is only dependent on human response time behavior, and therefore, only on workload behavior, we derive the expected reward function for the workload POMDP model as $\mathcal{R}_W: \set{W} \times \set{W} \times \mathcal{A} \rightarrow \mathbb{R}$  by calculating $\E{R|s_W,s_W',a}$, where random variable $R$ is the reward and $\E{\sbt}$ is the expected value of $\sbt$. 

Given the reward function $\mathcal{R}_{RT} \left(o_{RT} \right) = -o_{RT}$ and workload emission probability function $\mathcal{E}_{W}(o_{RT}|s_W)$ as represented in Figure~\ref{fig_emiss_Workload}, an equivalent expected reward function in the form $\mathcal{R}_W(s_W,s_W',a)$ is 
\begin{align} \label{eq_workload_rewards}
    \mathcal{R}_W(s_W,s_W',[a_{S_A},a_E,a_\tau]) = -(\mu_{s_W'} + \tau_{s_W'}),
\end{align}
where $\mu_{s_W'}$ and $\tau_{s_W'}$ are the parameters of the ex-Gaussian distribution corresponding to state $s_W'\in\set{W}$. We define the total reward function $\mathcal{R}$ for human trust-workload behavior as a convex combination of \eqref{eq_trust_rewards} and \eqref{eq_workload_rewards} with weight $\zeta$ as 
\begin{align}
    \mathcal{R}=\zeta\mathcal{R}_T+(1-\zeta)\mathcal{R}_W \enskip .
\end{align} 
As the weight $\zeta$ increases, more importance is given to the trust reward than the workload reward. For situations in which a correct decision is more important than a faster response time, a higher value of $\zeta$ should be used. Lastly, the discount factor $\gamma$ is selected based on the number of trials per mission in our study, i.e., $N=15$. We select the discount factor $\gamma$ such that the reward of the $15^{th}$ trial has a weight of $e^{-1}$; such a value of $\gamma$ can be approximated as
\begin{align}
\gamma= \frac{N}{N+1} = 0.9375 \enskip .
\end{align}
With the defined reward function and discount factor, we calculate the control policy for the POMDP model using the Q-MDP method as described in the next section.

\subsection{POMDP Control Policy} \label{sec_cont_solution} 
Using the reward function defined in the previous section, we determine the optimal control policy for updating the decision-aid's transparency by solving the combined trust-workload model to maximize the reward function defined in the previous section. Although it is possible to obtain the exact solution of the optimization through dynamic programming using value iteration, the time complexity increases exponentially with the cardinality of the action and observation spaces. Since a real-world scenario can involve a much larger set of actions and observations, obtaining the exact optimal solution may be intractable. Therefore, we adopt an approximate greedy approach called the Q-MDP method \cite{cassandra1994acting} to obtain a near optimal transparency control policy. The Q-MDP method solves the underlying MDP by ignoring the observation probability function to obtain the Q-function $Q_{\text{MDP}}:\mathcal{S}\times \mathcal{A} \rightarrow \mathbb{R}$. $Q_{\text{MDP}}(s,a)$ is the optimal expected reward given an action $a$ is taken at the current state $s$. Then, using the belief state $b(s)$, which can be iteratively calculated as
\begin{align}
b'(s') = \pr(s'|o,a,b(s)) = \frac{\pr(o|s',a) \sum\limits_{s\in \mathcal{S}}\pr(s'|s,a)b(s)}{\sum\limits_{s'\in \mathcal{S}}\pr(o|s',a) \sum\limits_{s\in \mathcal{S}}\pr(s'|s,a)b(s)} \enspace ,
\end{align}
the optimal action $a^*$ is chosen as 
\begin{align}
a^* = \argmax_a \sum_{s \in \mathcal{S}} b(s)Q_{\text{MDP}}(s,a)\enskip .
\end{align}
Essentially, the Q-MDP method approximates the optimal solution by assuming that the POMDP becomes completely observable after the next action. In order to solve the POMDP using the Q-MDP method, we calculate the Q-function of the underlying MDP using value iteration  \cite{puterman2014markov}. Nevertheless, as with any other method, the solution assumes that the decision-aid system can take any action $a\in \mathcal{A}$ in the future. But, in our model, only transparency $a_\tau$ is a controllable action; the other actions---recommendation $a_{S_A}$ and experience $a_E$---depend on the context and cannot be explicitly controlled by the policy. To account for these ``uncontrollable'' actions while solving for the control policy in the Q-MDP method, we calculate an expected Q-function of the form $Q^\tau:\mathcal{S}\times \tau \rightarrow \mathbb{R}$. This intermediate  Q-function is only dependent on the controllable actions and considers the probabilities of the uncontrollable actions. Finally, we iteratively solve \eqref{eq_cont_valueiter} until convergence is achieved to obtain $Q_{\text{MDP}}(s,a)$.
\begin{align} \label{eq_cont_valueiter}
\begin{split}
Q_{\text{MDP}}(s,a) &= \sum\limits_{s'\in \mathcal{S}}\mathcal{T}(s'|s,a)\left( \mathcal{R}(s'|s,a) + \gamma V(s') \right) \\
Q^\tau(s,\tau) &= \sum\limits_{a_{S_A}\in S_A, a_E \in E} \pr(a_{S_A},a_E)Q_{\text{MDP}}(s,[a_{S_A},a_E,a_\tau]) \\
V(s) &= \max_\tau Q^\tau(s,\tau)
\end{split}
\end{align}
Furthermore, $\pr(a_{S_A},a_E)=\pr(a_{S_A})\pr(a_E)$ because the present recommendation $a_{S_A}$ and experience $a_E$ due to the reliability of the last recommendation are independent. Therefore, $\pr(a_{S_A})$ and $\pr(a_E)$ are calculated as
\begin{align}
\begin{split}
\pr({S_A}^-) &= \beta d + (1-\alpha)(1-d) \enskip , \\
\pr({S_A}^+) &= 1-\pr(S_A^-) \enskip , \\
\pr(E^-) &= \alpha (1-d) + \beta d \enskip , \\
\pr(E^+) &= 1-\pr(E^-) \enskip .
\end{split}
\end{align}
For our human subject study, $d=0.5$, $\alpha =0.3$, and $\beta =0.3$. For implementation, once $a_{S_A}$ and $a_E$ are known in a trial, near-optimal transparency $a_\tau^*$ can be determined as
\begin{align}
a_\tau^* = \argmax_{a_\tau} \sum_{s \in \mathcal{S}} b(s)Q_{\text{MDP}}(s,[a_{S_A},a_E,a_\tau])\enskip .
\end{align}
We calculate the total reward function $\mathcal{R}$ and the corresponding control policy for three values of reward weights $\zeta= 0.50$, $\zeta= 0.91$, and $\zeta= 0.95$.

\begin{figure}
\centering
\subfigure[\label{fig_cont_sol_50_1}]{\includegraphics[width=0.23\textwidth]{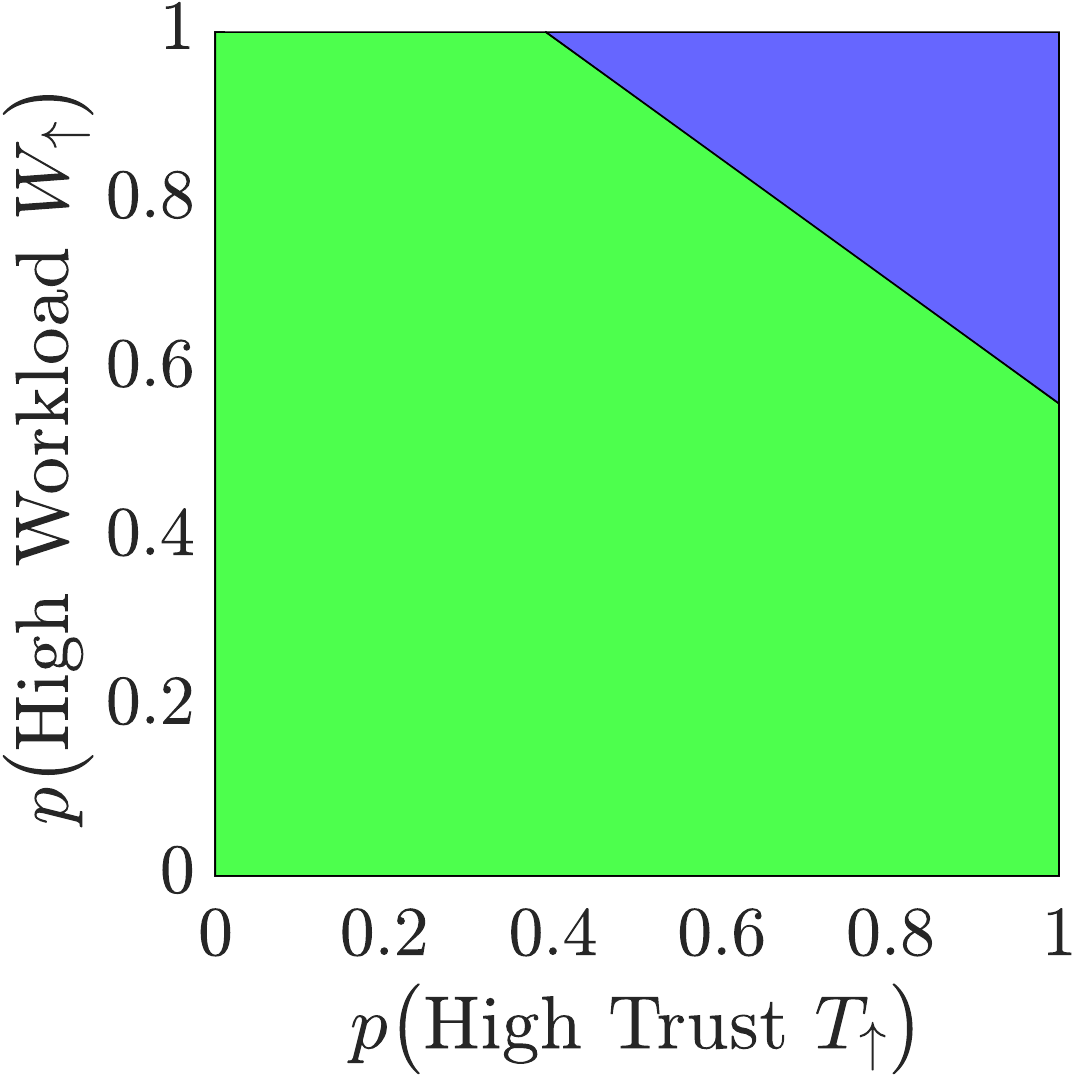}}
~
\subfigure[\label{fig_cont_sol_50_2}]{\includegraphics[width=0.23\textwidth]{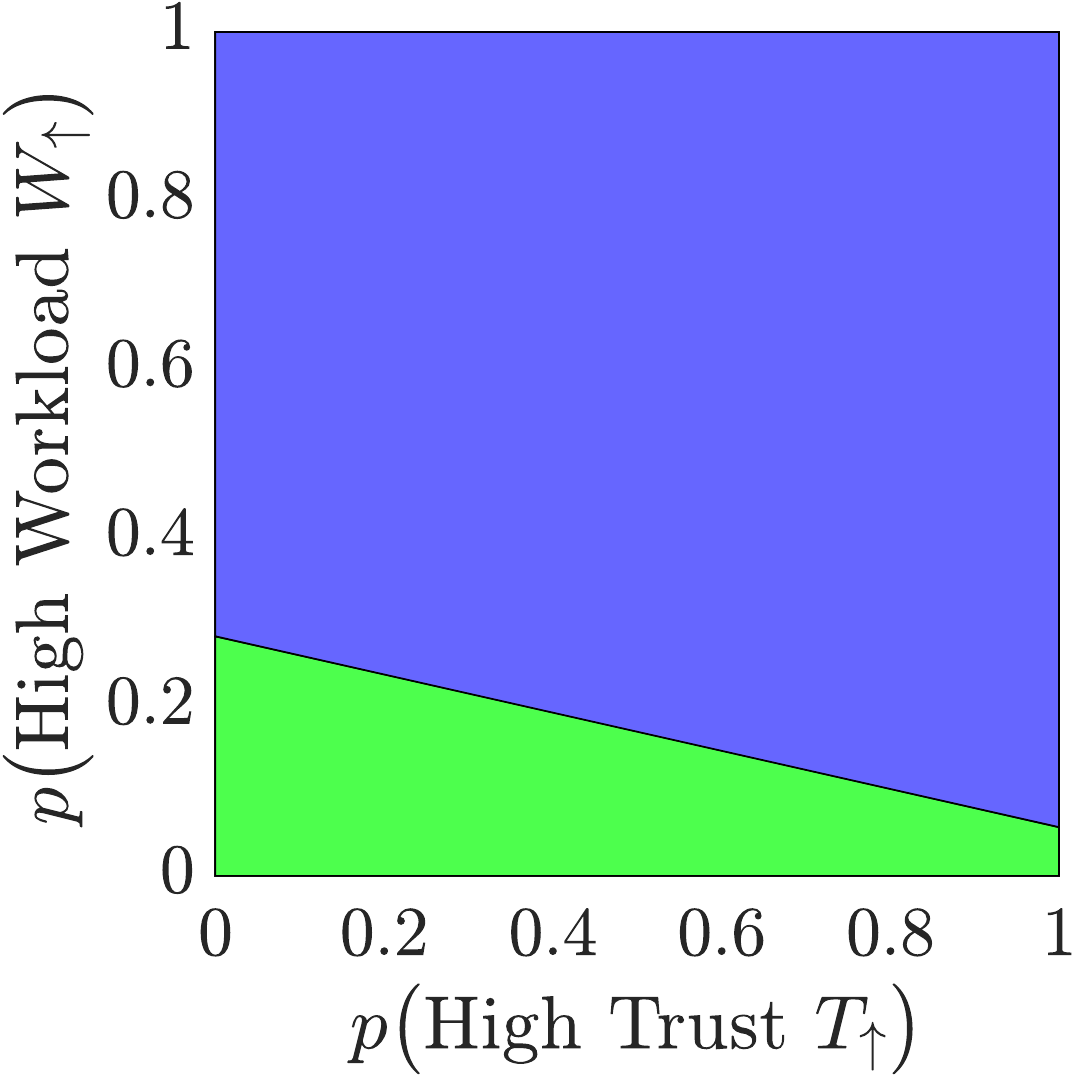}}

\subfigure[\label{fig_cont_sol_50_3}]{\includegraphics[width=0.23\textwidth]{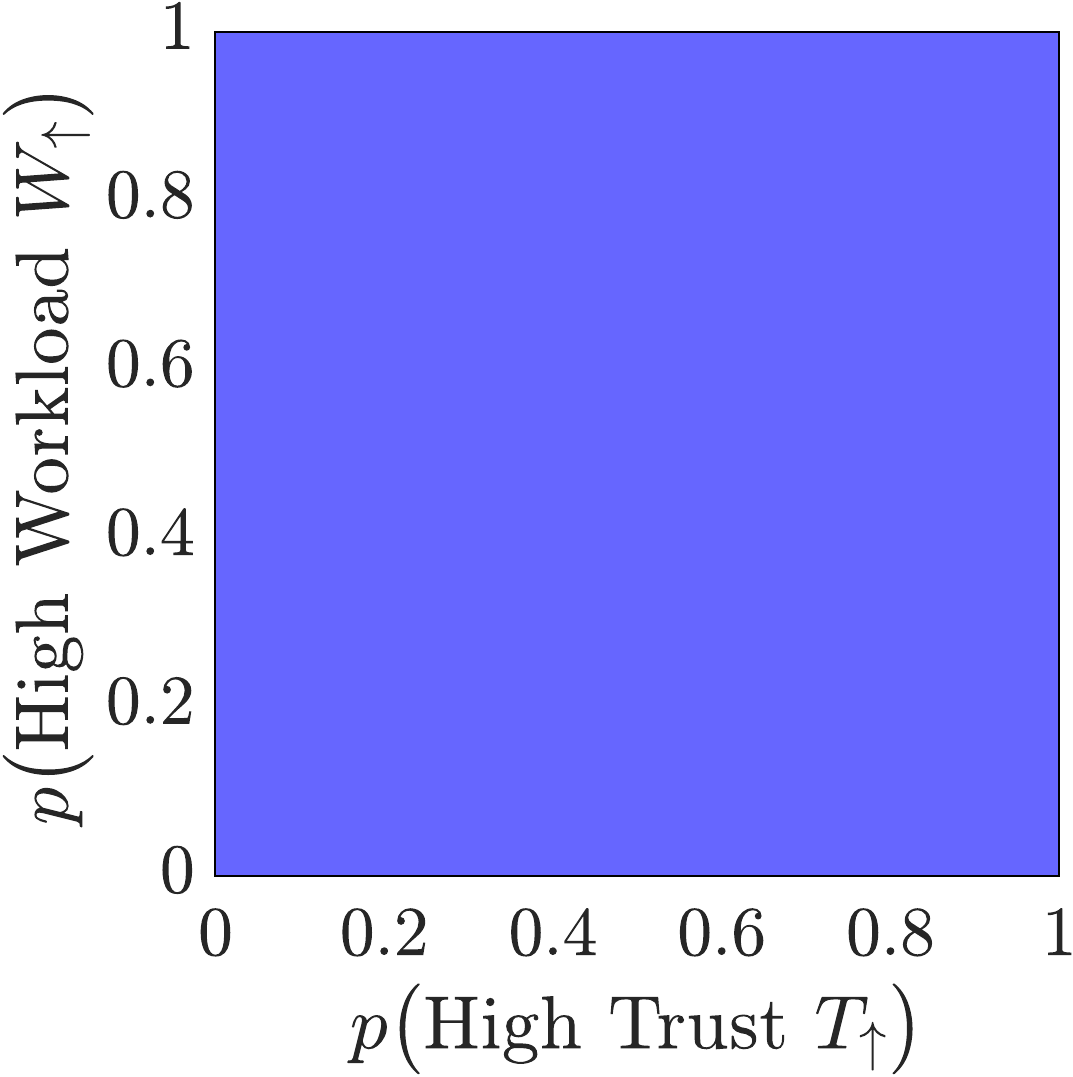}}
~
\subfigure[\label{fig_cont_sol_50_4}]{\includegraphics[width=0.23\textwidth]{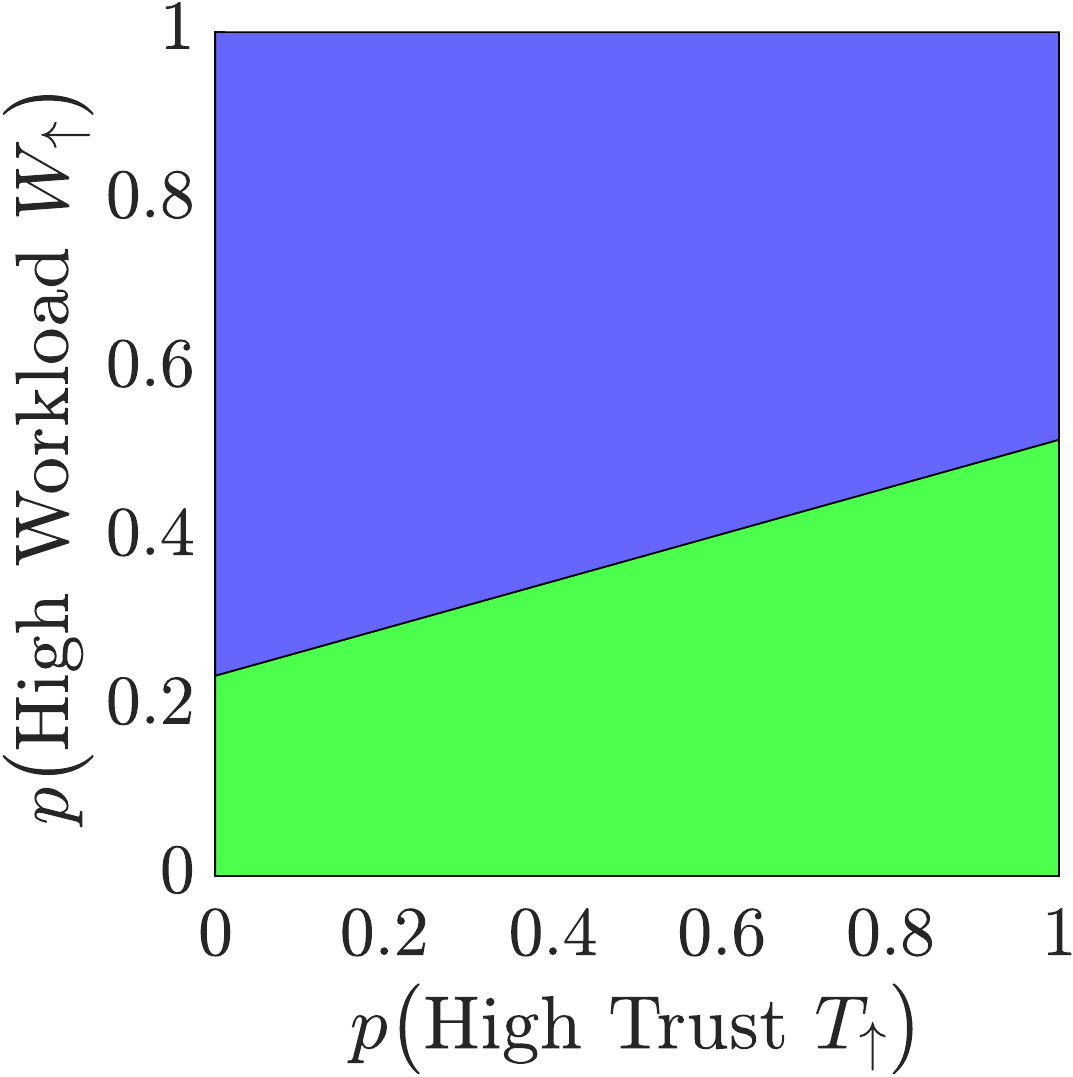}}

\subfigure{\includegraphics[width=0.225\textwidth]{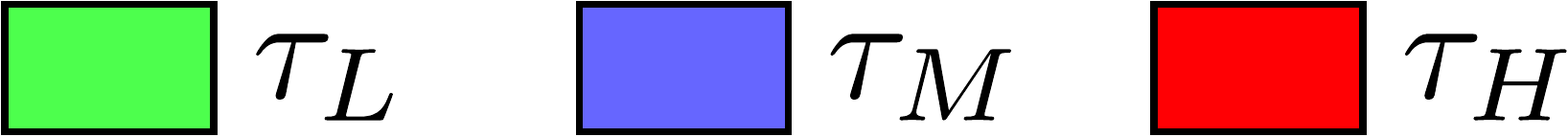}}
\caption{Closed-loop control policy corresponding to the reward function with $\zeta=0.50$. In this case, the reward function gives equal importance to the decision and response time rewards. Subfigure (a) corresponds to $a_{S_A}=S_A^-, a_E=E^-$, (b) corresponds to $a_{S_A}=S_A^-, a_E=E^+$, (c) corresponds to $a_{S_A}=S_A^+, a_E=E^-$, and (d) corresponds to $a_{S_A}=S_A^+, a_E=E^+$. When $\zeta=0.50$, high transparency is never adopted because it would result in a significant increase in response time.} 
\label{fig_cont_sol_50}
\end{figure}

\begin{figure}
\centering
\subfigure[\label{fig_cont_sol_91_1}]{\includegraphics[width=0.23\textwidth]{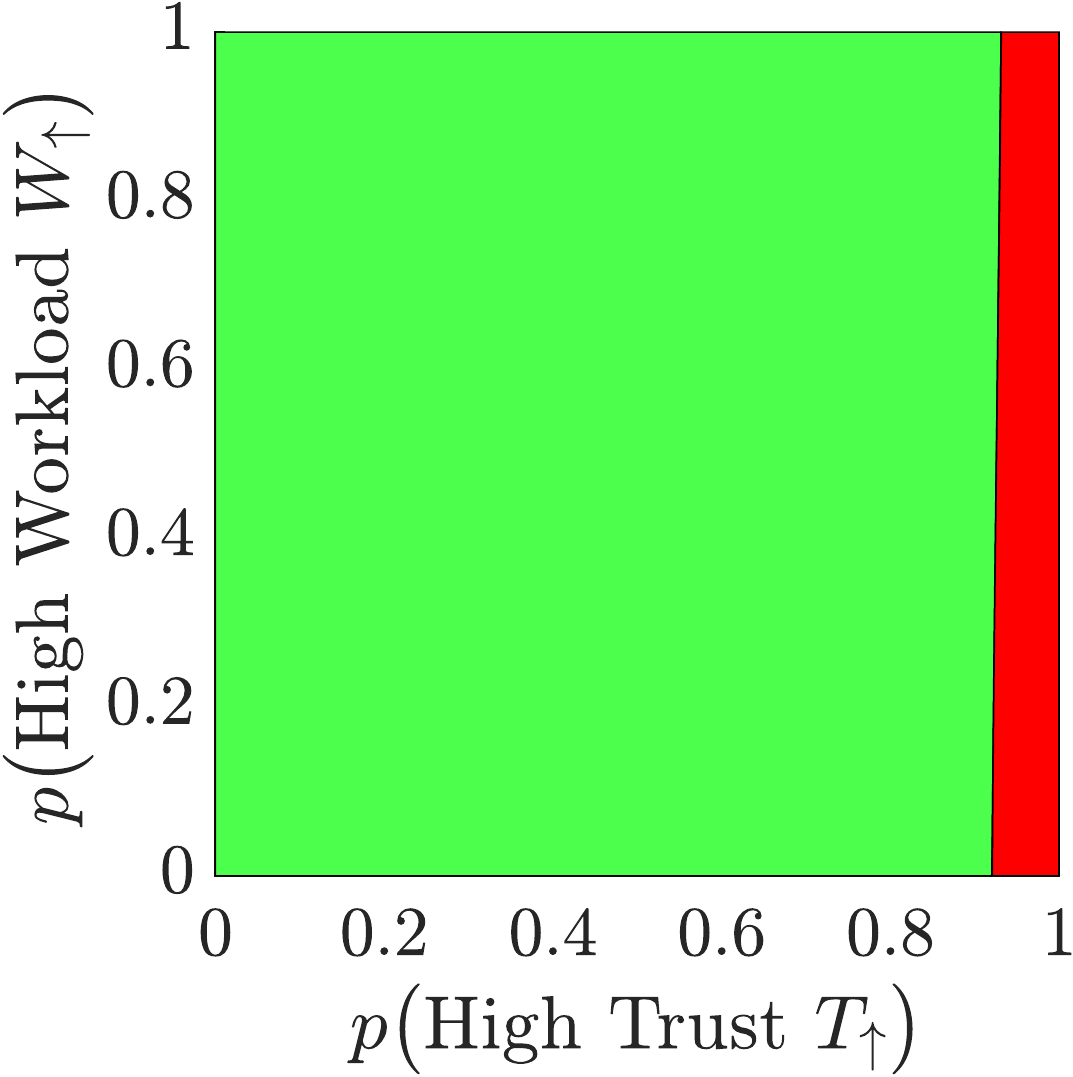}}
~
\subfigure[\label{fig_cont_sol_91_2}]{\includegraphics[width=0.23\textwidth]{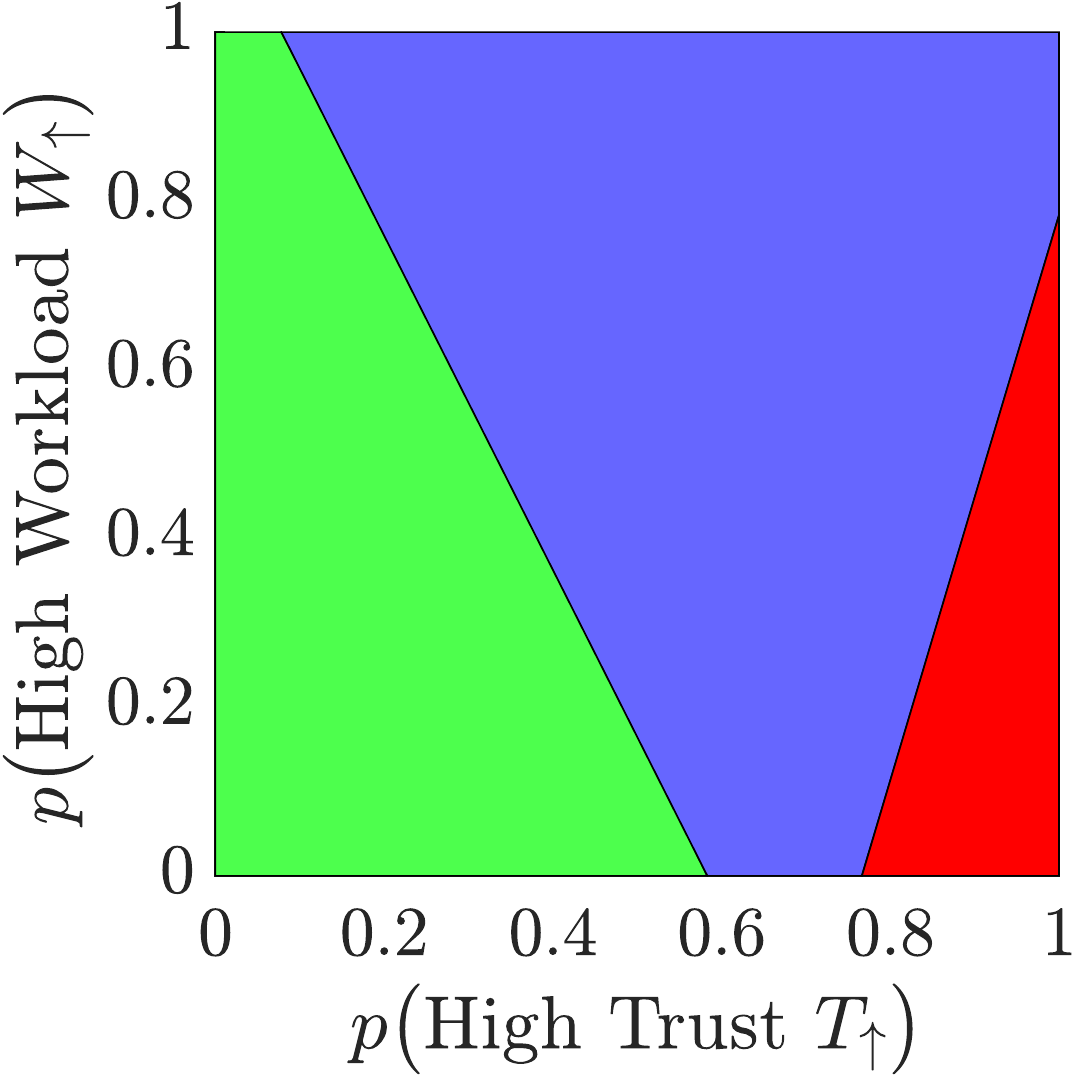}}

\subfigure[\label{fig_cont_sol_91_3}]{\includegraphics[width=0.23\textwidth]{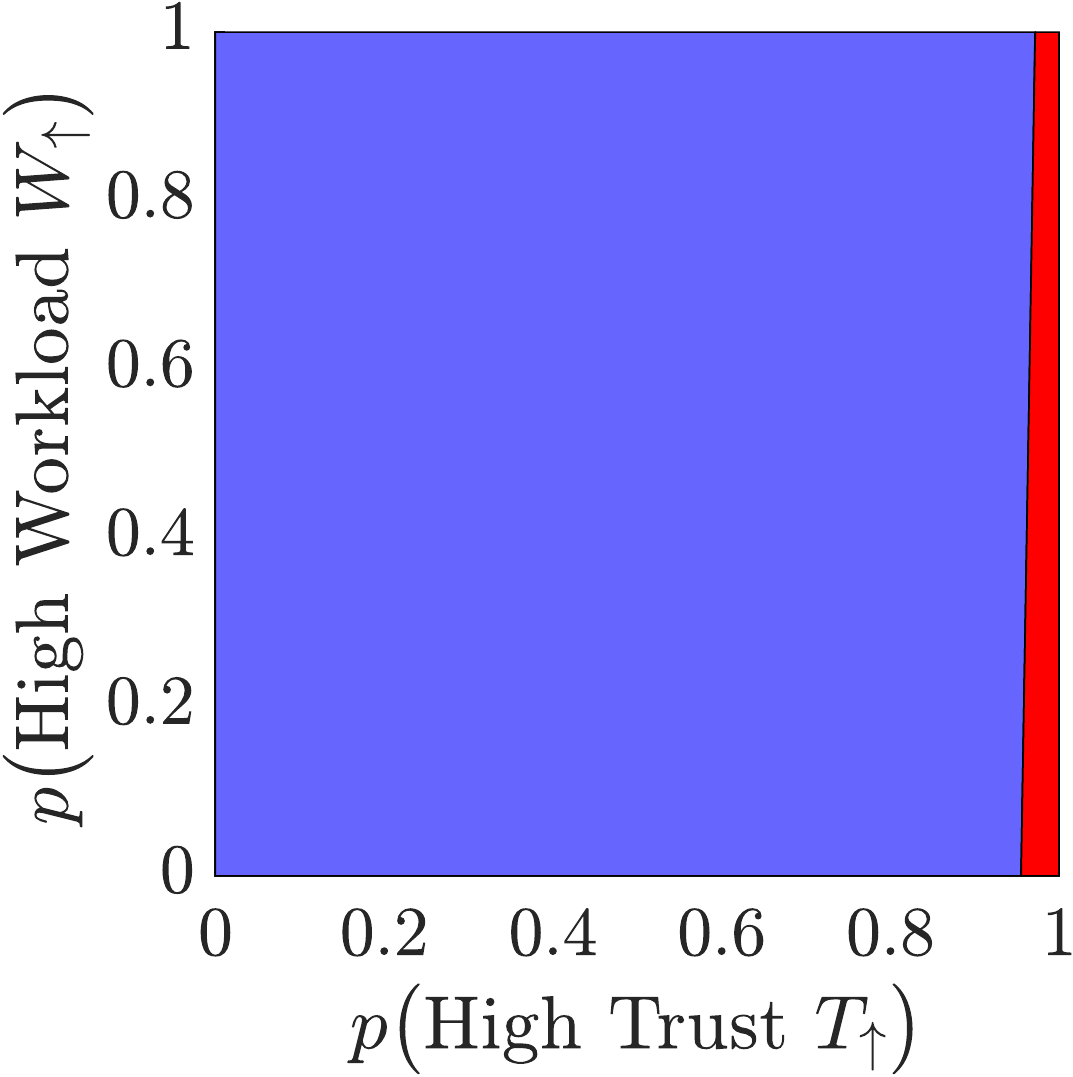}}
~
\subfigure[\label{fig_cont_sol_91_4}]{\includegraphics[width=0.23\textwidth]{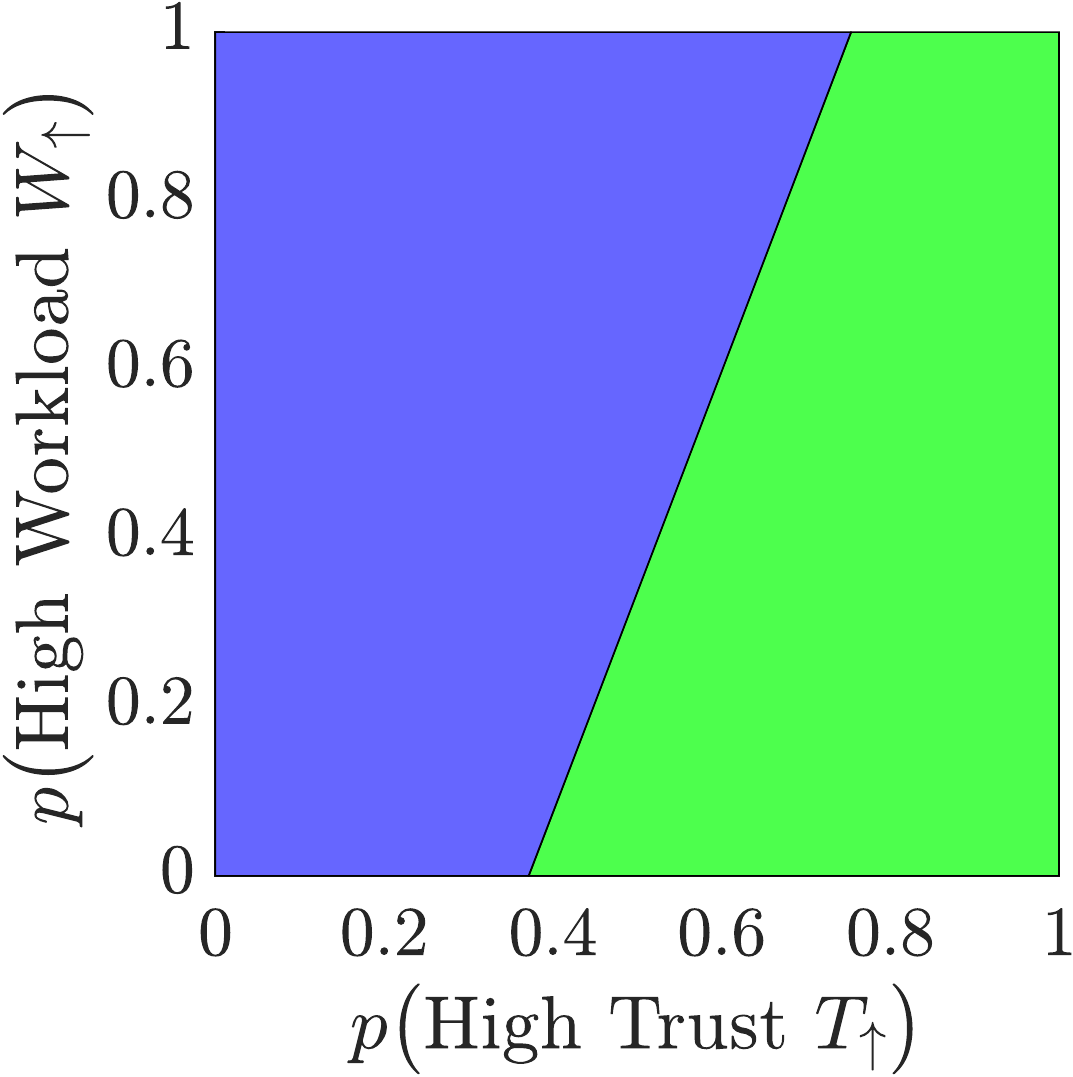}}

\subfigure{\includegraphics[width=0.225\textwidth]{figures/figure_9_10_11_policy_legend}}
\caption{The closed-loop control policy corresponding to the reward function with $\zeta=0.91$. In this case, higher importance is given to the decision rewards as compared to the response time rewards. Subfigure (a) corresponds to $a_{S_A}=S_A^-, a_E=E^-$, (b) corresponds to $a_{S_A}=S_A^-, a_E=E^+$, (c) corresponds to $a_{S_A}=S_A^+, a_E=E^-$, and (d) corresponds to $a_{S_A}=S_A^+, a_E=E^+$. This control policy adopts high transparency for very high probabilities of High Trust to reduce the number of incorrect decisions the human may make due to their over-trust in the decision-aid system.}
\label{fig_cont_sol_91}
\end{figure}

\begin{figure}
\centering
\subfigure[\label{fig_cont_sol_95_1}]{\includegraphics[width=0.23\textwidth]{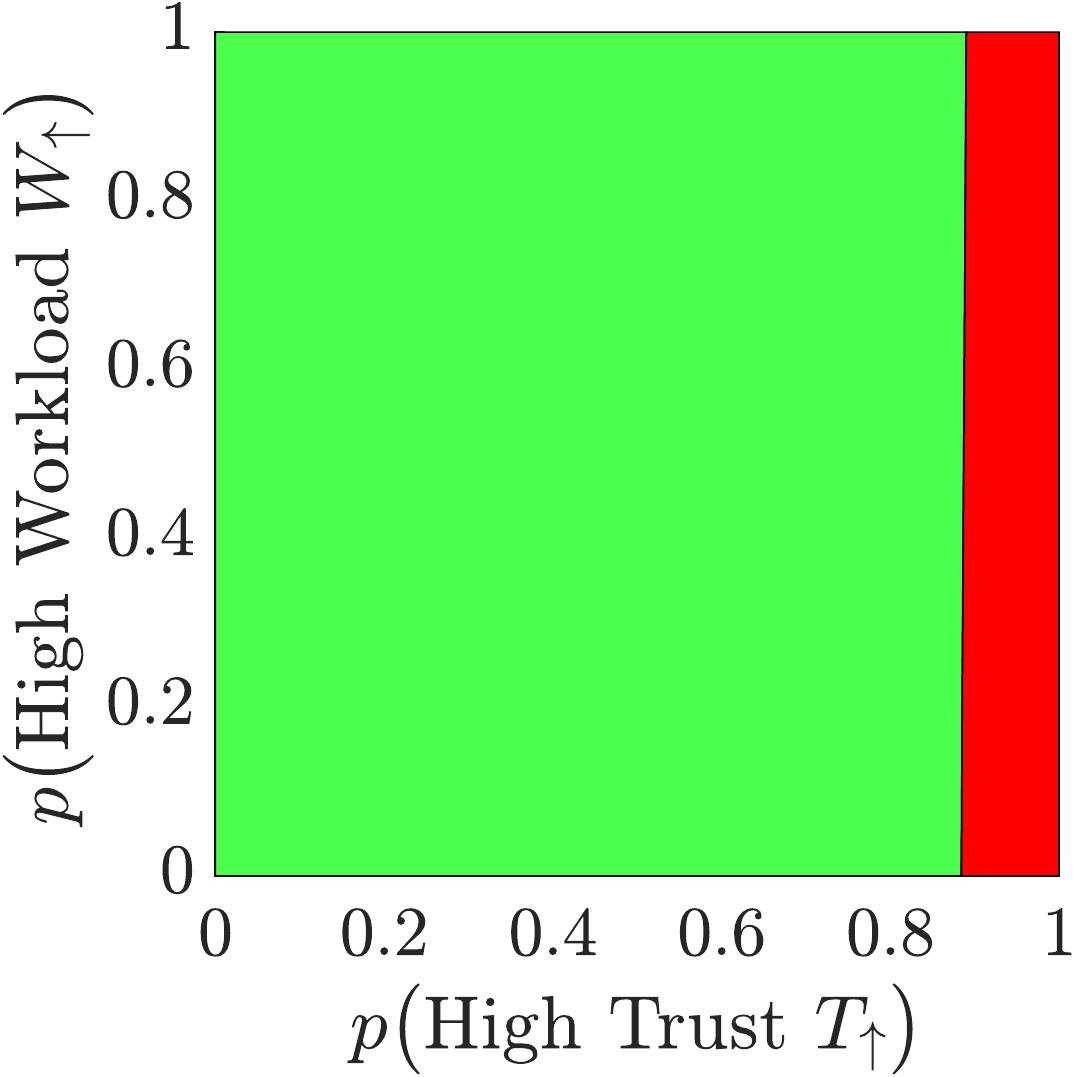}}
~
\subfigure[\label{fig_cont_sol_95_2}]{\includegraphics[width=0.23\textwidth]{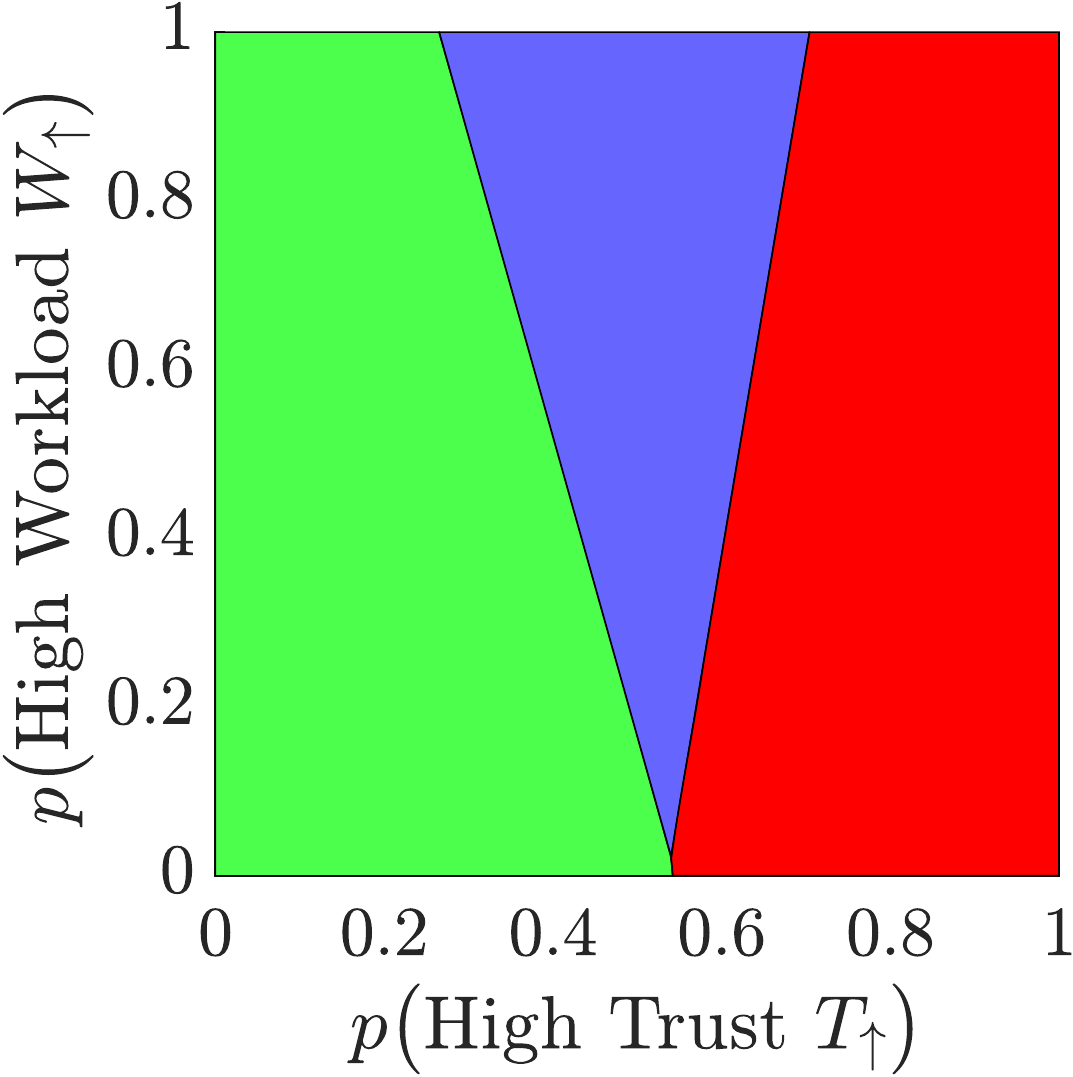}}

\subfigure[\label{fig_cont_sol_95_3}]{\includegraphics[width=0.23\textwidth]{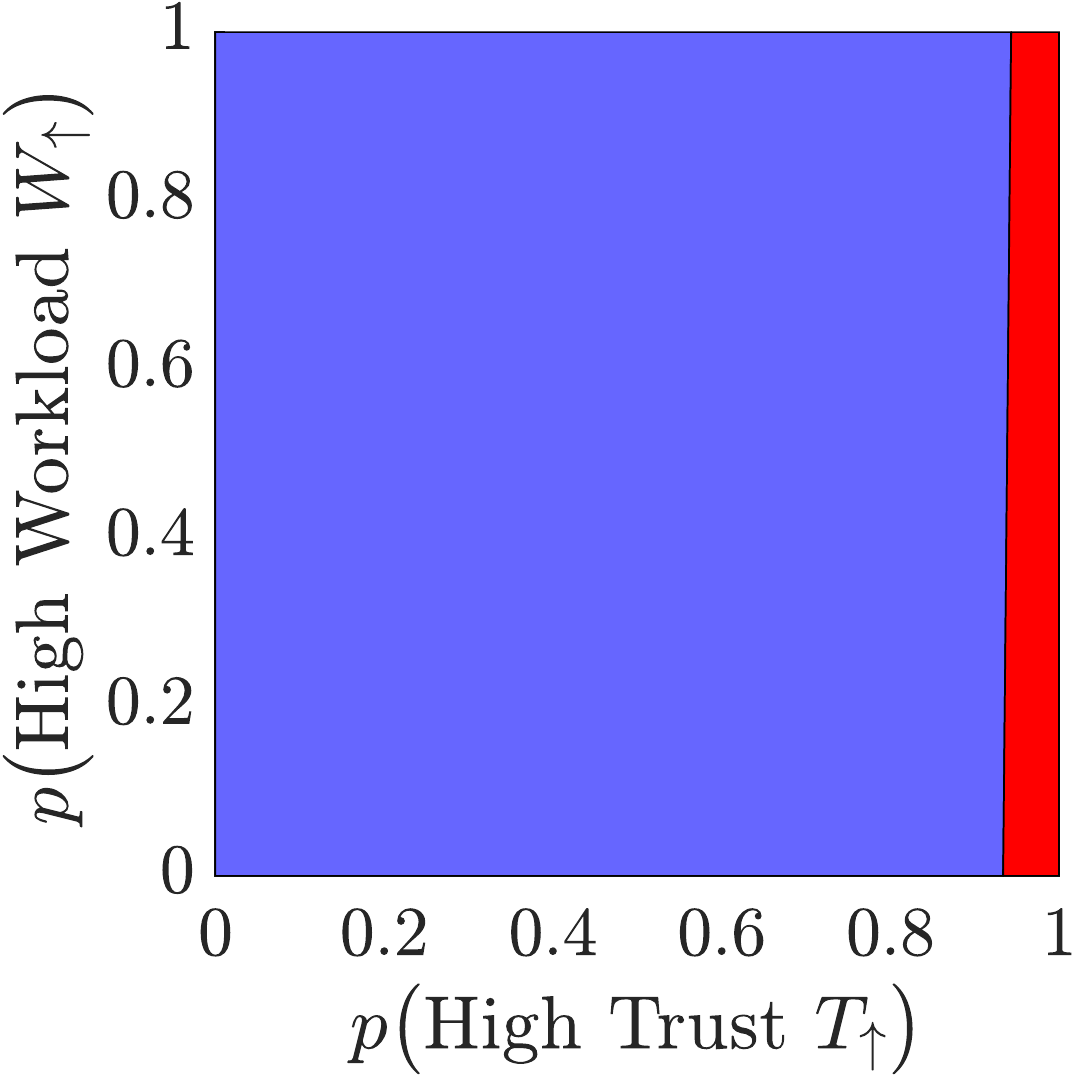}}
~
\subfigure[\label{fig_cont_sol_95_4}]{\includegraphics[width=0.23\textwidth]{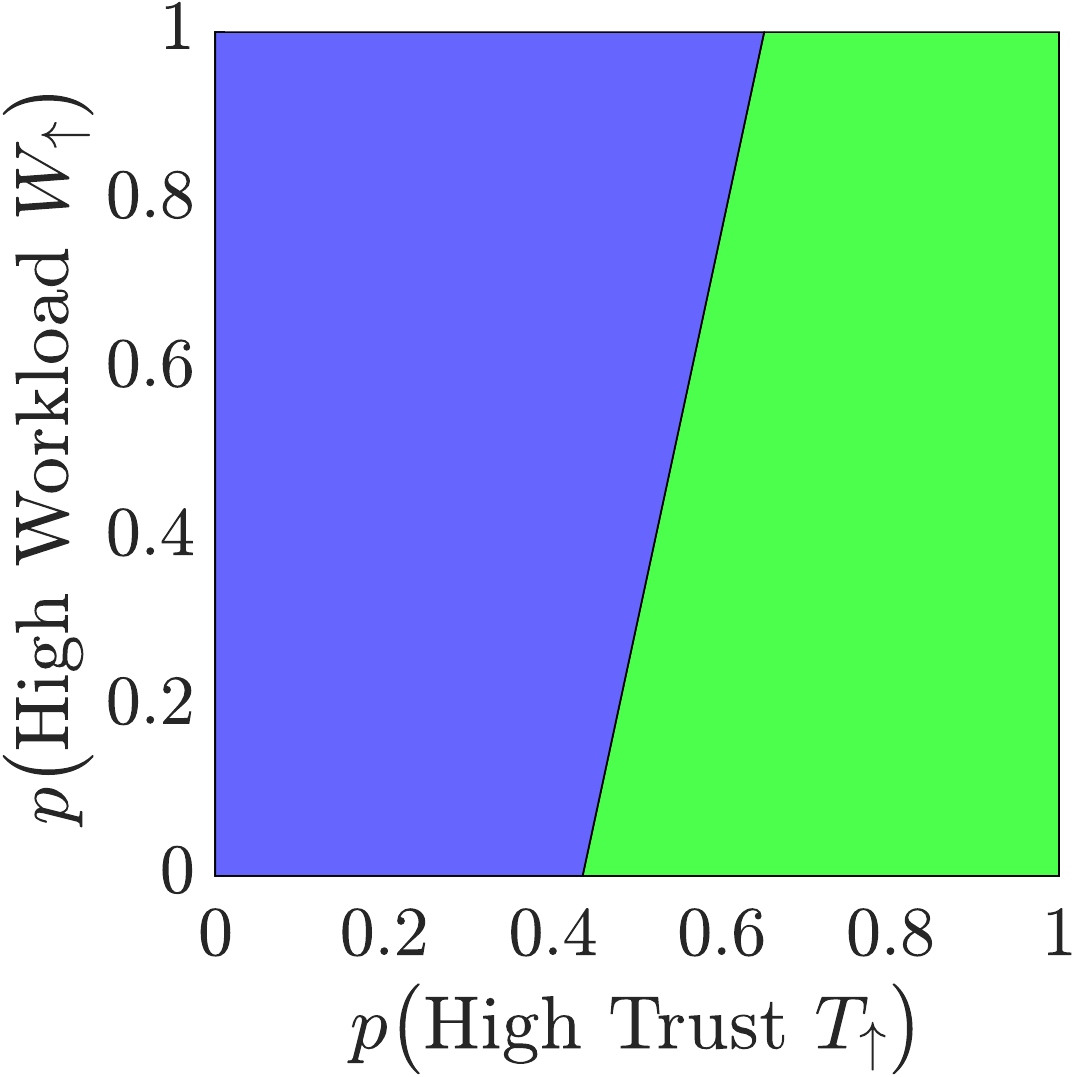}}

\subfigure{\includegraphics[width=0.225\textwidth]{figures/figure_9_10_11_policy_legend}}
\caption{The closed-loop control policy corresponding to the reward function with $\zeta=0.95$. In this case, a very high importance is given to the decision rewards as compared to the response time rewards. Subfigure (a) corresponds to $a_{S_A}=S_A^-, a_E=E^-$, (b) corresponds to $a_{S_A}=S_A^-, a_E=E^+$, (c) corresponds to $a_{S_A}=S_A^+, a_E=E^-$, and (d) corresponds to $a_{S_A}=S_A^+, a_E=E^+$.  This control policy again adopts high transparency for high probabilities of High Trust to reduce the number of incorrect decisions the human may make due to their over-trust in the decision-aid.}
\label{fig_cont_sol_95}
\end{figure}

The control policies corresponding to each of the reward weights are depicted in Figures~\ref{fig_cont_sol_50}, \ref{fig_cont_sol_91}, and \ref{fig_cont_sol_95}, respectively. We first consider the case with $\zeta= 0.50$ shown in Figure~\ref{fig_cont_sol_50}.  Here, the reward function gives equal importance to the decision and response time rewards. Each of the four figures represents the optimal choice of transparency based on the estimated probability of High Trust $\Thigh$ and High Workload $W_\uparrow$ for a given recommendation $a_{S_A}$ and experience $a_E$. We first consider the case when the recommendation suggests Light Armor $a_{S_A} = S_A^-$ as shown in  Figures~\ref{fig_cont_sol_50_1} and \ref{fig_cont_sol_50_2}. This case represents a high risk situation for over-trust because an incorrect human decision of complying with the recommendation can lead to the human using Light Armor in the presence of gunmen, resulting in injury and an extra penalty of 20 seconds. The control policy adopts medium transparency $\tau_M$ when the probabilities of High Trust and High Workload are high. Medium transparency can help the human to make a more informed decision than low transparency, thereby avoiding over-trust in these cases when the human's trust is too high. Also, if the human's experience was reliable from the last trial ($a_E = E^+$), the chance of over-trust is higher. In this case, medium transparency is adopted at even lower probabilities of High Trust as seen in Figure~\ref{fig_cont_sol_50_2}. Furthermore, as seen in Figure~\ref{fig_transition_Workload}, medium transparency is best at transitioning a human from a state of High Workload to a state of Low Workload. Therefore, medium transparency will help to reduce the expected response time when the probability of High Workload is high. 

For the case when the decision-aid recommends Heavy Armor $a_{S_A} = S_A^+$, shown in  Figures~\ref{fig_cont_sol_50_3} and \ref{fig_cont_sol_50_4}, the situation risk is low given that an incorrect compliance only leads to an extra 4 seconds of search time. Therefore, the control policy always aims to increase trust in this case. Since medium transparency has the highest probability of causing a transition from Low Trust to High Trust (Figure~\ref{fig_Trust_trans_21}) and \ref{fig_Trust_trans_22}), the control policy adopts medium transparency when the probability of High Trust is low. When the probability of High Trust is high and the human's prior experience with the decision aid was Faulty, medium transparency has a higher probability of maintaining a high trust level as compared to low transparency (Figure~\ref{fig_Trust_trans_21}); therefore, the control policy adopts medium transparency in this case (Figures~\ref{fig_cont_sol_50_3}). Note that high transparency is not adopted by the control policy in this case due to the large response time penalty associated with high transparency. When the human's prior experience with the decision aid was reliable, low transparency has the highest probability of maintaining high trust level (Figure~\ref{fig_Trust_trans_22}); therefore, low transparency is preferred with low levels of workload (Figures~\ref{fig_cont_sol_50_4}). Moreover, medium transparency is adopted when the probability of High Workload is high as discussed above. In general, medium transparency dominates the control policy for $\zeta=0.50$ because in most cases it provides a good trade-off between trust calibration based on informed decision-making and increased workload.  

For the cases with $\zeta = 0.91$ and  $\zeta = 0.95$, higher importance is given to the decision rewards as compared to the response time rewards. In these cases, as represented in Figure~\ref{fig_cont_sol_91} and \ref{fig_cont_sol_95}, we observe that the control policies adopt high transparency for a very high probability of High Trust; this ensures that the human does not over-trust the automation and instead makes the most informed decision possible. With higher values of $\zeta$, the use of high transparency is further increased since the associated weight for the response time reward is significantly reduced.
In the next section, these control policies are implemented to dynamically vary transparency based on the participant's current trust and workload estimates in a reconnaissance mission study.

\section{Validation and Results} \label{sec_result}

To experimentally validate the performance of the proposed control policies represented in Figures~\ref{fig_cont_sol_50}, \ref{fig_cont_sol_91}, and \ref{fig_cont_sol_95}, we conducted two human subject studies. These experiments were identical to the one used to collect open-loop data for each transparency but with transparency controlled using the control policies based on human trust and workload estimation.

\emph{Stimuli and Procedure: }
Two within-subjects studies were performed in which participants were asked to interact with a simulation of three reconnaissance missions as described in the earlier study description. However, instead of fixed transparency in each mission, the transparency was controlled based on a feedback control policy for some missions. For the first study, high transparency was always used in one of the three missions, and in the other two missions, the transparency was dynamically varied based on control policies corresponding to $\zeta=0.50$ (Figure~\ref{fig_cont_sol_50}) and $\zeta=0.95$ (Figure~\ref{fig_cont_sol_95}), respectively. In the second study, the transparency was dynamically varied  based on the control policy corresponding to $\zeta=0.91$ (Figure~\ref{fig_cont_sol_91}) in one mission, and the other two missions used fixed medium transparency and fixed high transparency, respectively. Three missions in each study ensured that the studies were short enough to avoid participant fatigue and were consistent in structure with the study used to collect open-loop data. Moreover, the order of missions was again randomized across participants to reduce ordering effects~\cite{shaughnessy2012research}.

\emph{Participants: } 
One hundred and twenty participants for the first study, and one hundred and four participants for the second study, participated in and completed the study online. They were recruited using Amazon Mechanical Turk \cite{amazon2005amazon} with the same criteria used for the earlier study. To account for participants who were not sufficiently engaged in the study, we filtered data that had any response time higher than 40.45 seconds. As a result, 20 outlying participants were removed from the dataset for the first study, and 7 outlying participants were removed from the dataset for the second study, leading to a remaining 100 and 97 participants, respectively.

\begin{figure}
\centering
\includegraphics[width=0.67\textwidth]{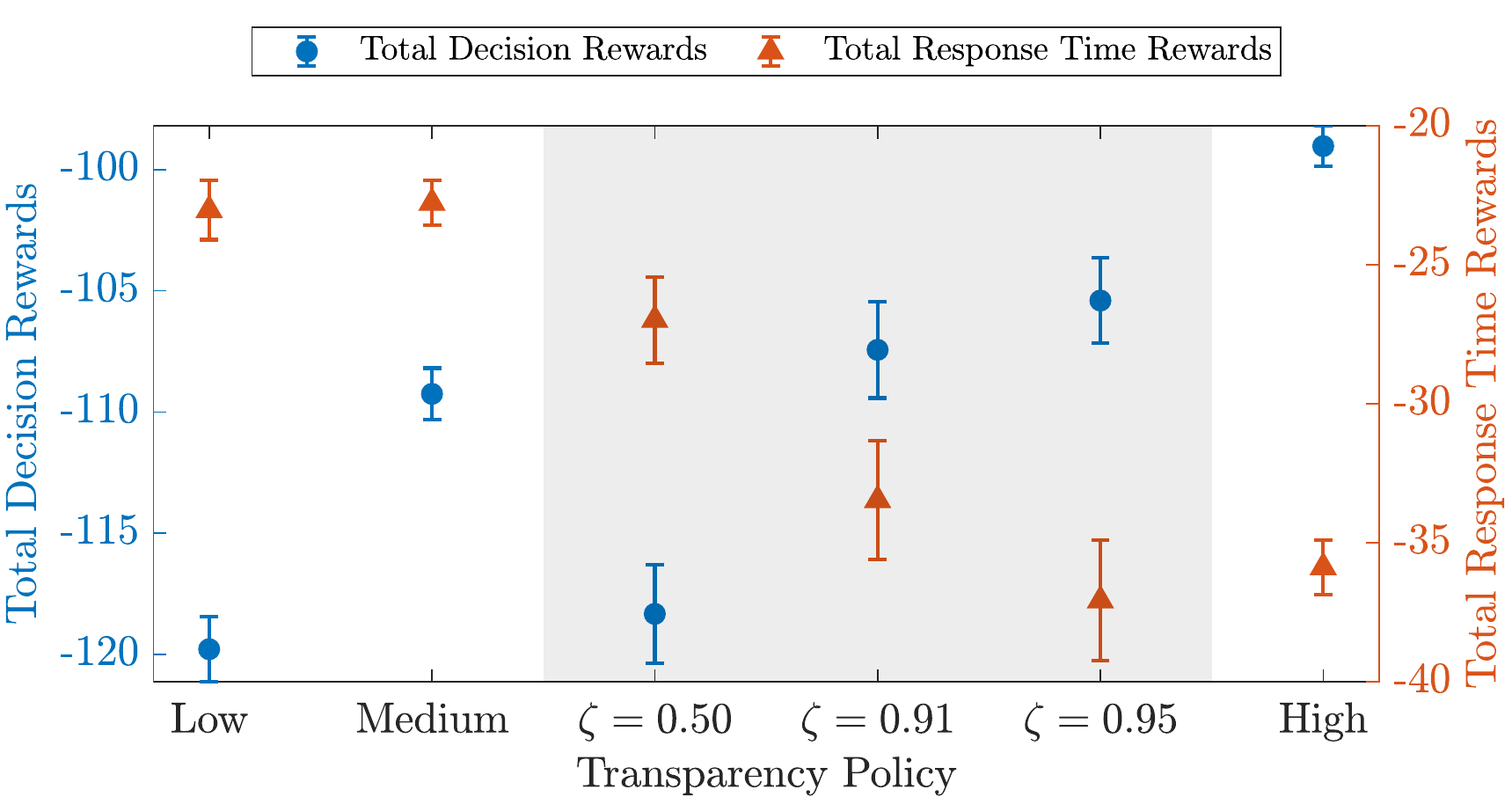}
\caption{Effect of the proposed control policies on the total decision and total response time rewards. Error bars represent the standard error of the mean across participants. The closed-loop control policies are highlighted in gray. The performance of the closed-loop policies lies between that of high and low transparency in terms of both reward metrics. With higher values of the reward weight $\zeta$, the performance of the closed-loop policy is more similar to that of high transparency. Depending on the requirements of the context, $\zeta$ can be tuned to achieve the required trade-off between decision and response time performance.}
\label{fig_decision_and_rt_rewards}
\end{figure}

Using the collected human subject data from the two validation studies along with the open-loop study discussed earlier, we quantify and evaluate participants' performance for the dynamically varying transparency missions and the fixed transparency mission. We compare two metrics: total decision reward and total response time reward for each type of mission. We use linear mixed effects analysis and  likelihood ratio tests to determine whether the use of trust-workload behavior-based feedback had any significant effect on these metrics. We used the statistical computing language {R} \cite{rcoreteam2018r} and \emph{lme4} library \cite{bates2015fitting} to perform a linear mixed model approach to analyze the relationship between each of the metrics and the transparency policies. As a fixed effect, we used the transparency policy in the models. To account for variations in the metrics calculated for different participants, the models considered each individual as a random effect. P-values were obtained using likelihood ratio tests of the full model that includes the transparency policy as a fixed effect against the model that does not include the transparency policy. Figure~\ref{fig_decision_and_rt_rewards} shows the effect of the transparency policies (open-loop: Low, Medium, and High; closed-loop: $\zeta=0.50,0.91, \text{and } 0.95$) on the total decision rewards and on the total response time rewards across participants.

The total decision reward is defined as the sum of all decision rewards based on Table~\ref{tab_rewards} accrued by the participant in a mission. A likelihood ratio test using linear mixed effects models indicated that the transparency policies significantly affected total decision rewards ($\chi^2(6)=229.62, p\approx 0.0000$). The total response time reward is defined as the negative of the sum of all response times in seconds accrued by the participant in a mission. A likelihood ratio test indicated that the transparency policies significantly affected total response time rewards ($\chi^2(6)=230.07, p\approx 0.0000$). 

To analyze the performance and benefit of the closed-loop control policies, we compare them with the performance of the open-loop cases (that consider a static transparency). Furthermore, we analyze the effects of the reward weight $\zeta$ on the closed-loop performance. From Figure~\ref{fig_decision_and_rt_rewards}, considering open-loop fixed transparency policies, we see that high transparency has the best performance in terms of decision rewards, followed by medium and low transparency. However, low and medium transparency perform better in terms of response time rewards; this indicates a trade-off between the correctness of the human's decision versus the corresponding response time. Although the use of high transparency can result in the highest number of correct decisions, high response times indicate higher workload levels for the human. Furthermore, some time-critical contexts may favour fast response times in lieu of perfect decisions.

Given this trade-off, we see that the performance of our closed-loop policies lies between that of high and low transparency in terms of both reward metrics (see gray-highlighted region in Figure~\ref{fig_decision_and_rt_rewards}). We see that with higher values of the weight $\zeta$, the performance of the closed-loop policy is more similar to that of high transparency used all the time. The control policy corresponding to $\zeta=0.91$ performs better than the medium transparency in terms of decision rewards but has lower response time rewards. Therefore, depending on the requirements of the context, the proposed control policy enables the controls engineer to trade off between decision and response time performance.

The framework presented here provides a substantial step forward toward the development of quantitative dynamic models of human behavior and their use for implementing adaptive automation in human-machine interaction contexts. Nevertheless, it should be noted that the overall performance of the control policy could be improved by addressing a few limitations of the proposed trust-workload model. We assumed that human trust and workload behavior are conditionally independent to simplify the model structure and complexity. However, trust and workload may be coupled, and therefore changes in the trust state could directly impact the workload state and vice-versa. 
Moreover, while the closed-loop policy resulted in a sometimes rapid change of transparency between trials, the model parameters were estimated using data with only fixed transparency. Therefore, the model does not account for the effect of such changes on the human's trust and workload. It is possible that this effect is not negligible, and future work could consider retraining the model with this type of data. 

\section{Conclusions} \label{sec_conclusion}

Interactions between humans and automation can be improved by designing automation that can infer human behavior and respond accordingly. We developed a model of human trust and workload dynamics as they evolve during a human's interaction with a decision-aid system, and further designed and validated a model-based feedback control policy aimed at dynamically varying the automation's transparency to improve the overall performance of the human-machine team. The model, which was parameterized using human subject data, captured the effects of the decision-aid's recommendation, the human's previous experience with the automation, and automation transparency on the human's trust-workload behavior. The model is capable of estimating human trust and workload in real time using recursive belief-state estimates. Furthermore, experimental validation showed that the closed-loop control policies were successfully able to manage the human decision versus response time performance tradeoff based on a tuning parameter in the reward function. This framework provides a tractable methodology for using human behavior as a real-time feedback signal to optimize human-machine interactions through dynamic modeling and control. 

Although this work focused on a case study of a reconnaissance mission task, the model framework could be used in a variety of other decision-aid contexts, such as health recommender systems and other assistive robot applications, by retraining the model using new context-specific data. Future work could consider extensions of the framework to action automation by re-defining context-specific observations, actions, and reward function(s). Furthermore, since a computer-based simulated interface was used in the experiment, the ecological validity could be improved by testing the established framework in both immersive environments, for example, flight or driving simulators, as well as real-life settings.

\appendices
\section{What is a POMDP?}\label{sb_POMDP}

A Partially Observable Markov Decision Process (POMDP) is an extension of a Markov decision process (MDP) that accounts for partial observability through hidden states. It is similar in structure to the classic discrete-time state-space model as described in Table~\ref{tab_POMDP_sidebar} but with a discrete state-space.
Formally, a POMDP is a 7-tuple ($\mathcal{S}, \mathcal{A}, \mathcal{O}, \mathcal{T},  \mathcal{E}, \mathcal{R}, \gamma $) where $\mathcal{S}$ is a finite set of states, $\mathcal{A}$ is a finite set of actions, and $\mathcal{O}$ is a set of observations. The transition probability function $\mathcal{T}(s'|s,a)$ governs the transition from the current state $s$ to the next state $s'$ given the action $a$. The emission probability function $\mathcal{E}(o|s)$ governs the likelihood of observing $o$ given the process is in state $s$. Finally, the reward function $\mathcal{R}(s',s,a)$ and the discount factor $\gamma$ are used for finding an optimal control policy. A detailed description of MDPs and POMDPs can be found in \cite{sigaud2013markov}.
\begin{table}[hb]
\centering
\caption{Similarities between a Partially Observable Markov Decision Process (POMDP) and a discrete-time state-space model.}\label{tab_POMDP_sidebar}
\begin{tabular}{|l|c|c|}
\hline
    & POMDP & State-space model  \\\hline
States & $s\in \mathcal{S}$      &  $x\in \mathbb{R}_x$ \\
Actions/Inputs       & $a\in \mathcal{A}$ & $u\in \mathbb{R}_u$ \\
Observations/Outputs       & $o\in \mathcal{O}$ & $y\in \mathbb{R}_y$ \\
Transition function       & $\pr(s') = \mathcal{T}(s'|s,a)$ & $x_{t+1} = f(x_t,u_t)$ \\
Emission/Output function       & $\pr(o) = \mathcal{E}(o|s)$ & $ y_t  = g(x_t)$ \\
Reward/Cost function       & $\mathcal{R}(s',s,a)$ & $\mathcal{L}(x_t,u_t)$ \\
Optimal control policy       & $a^* = \argmax\limits_{a\in \mathcal{A}} \sum\limits_{t=0}^\infty \gamma^t \mathcal{R}(s^{t+1},s^t,a^t)$ & $u^* = \argmin\limits_{u\in \mathbb{R}_u} \sum\limits_{t=0}^\infty \mathcal{L}(x_t,u_t)$ \\ 
\hline
\end{tabular}
\end{table}

\section{Response time and its distribution}\label{sb_exGauss}
\begin{figure}[ht]
  \centering
  \includegraphics[width=.55\textwidth]{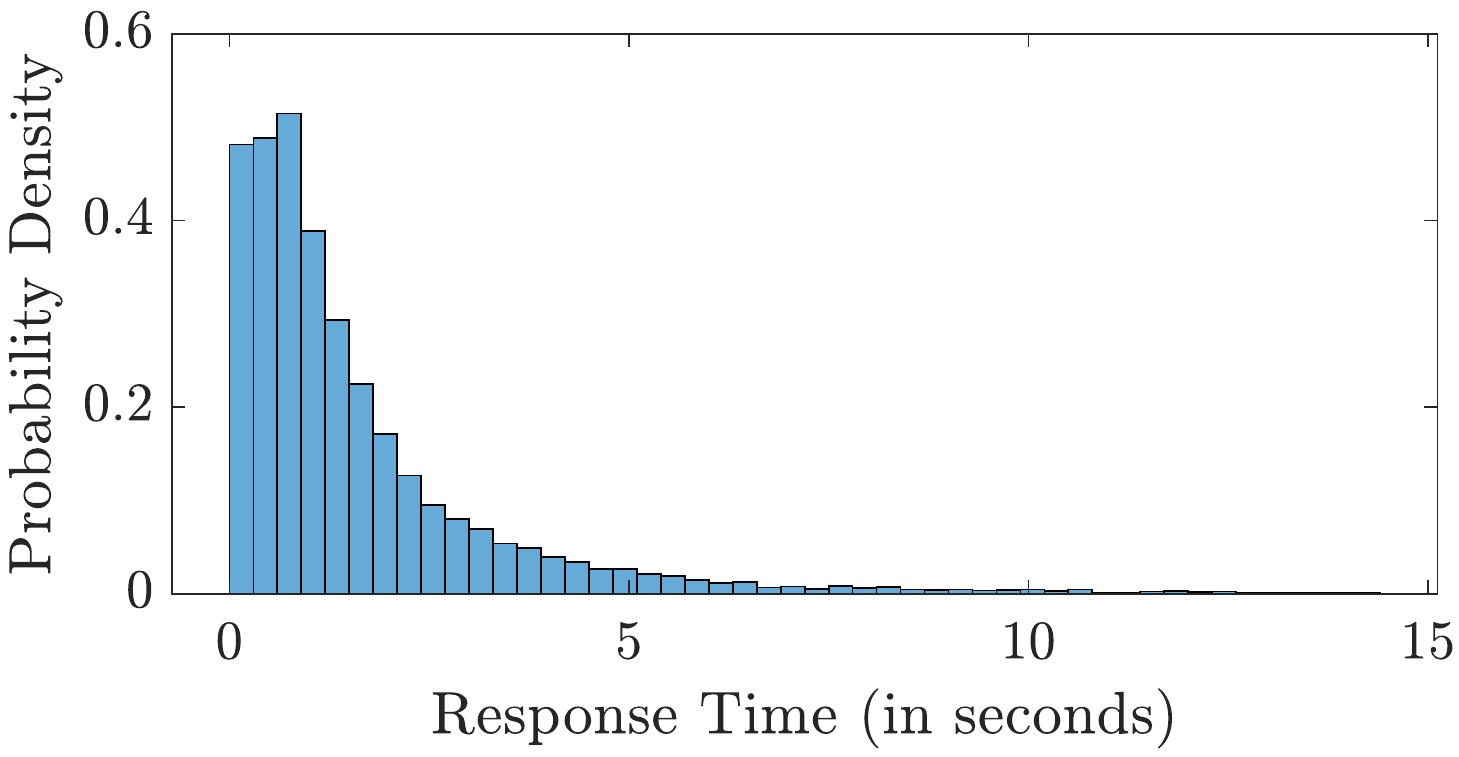}
  \caption{Empirical probability density function representing the response time $RT$ distribution for the aggregated human subject study data described in this article. $RT$ distributions are attributed with a positively skewed unimodal shape with a rapid rise on the left and a long positive tail on the right.}
  \label{fig_resptime_dist}
\end{figure}

Human response time $RT$ (also called reaction time or latency \cite{whelan2008effective}) is the time duration between the presentation of stimulus to a human and the human's response \cite{luce1986response}. Response time analysis has a long history in experimental psychology and still is used as a dominant dependent measure to identify the processes that affect the human's response. Statistically, $RT$ is often treated as a random variable because it typically varies between trials for the same human within a given context.  Furthermore, $RT$ distributions are attributed with a positively skewed unimodal shape (see Figure~\ref{fig_resptime_dist}) that cannot be effectively captured by only mean and variance \cite{jaskowski1983distribution}. Therefore, a $RT$ distribution cannot be modeled as a Gaussian distribution.

To describe a typical $RT$ distribution, standard distributions like gamma distribution and log-normal distribution have been used in the literature \cite{ratcliff1976retrieval, mcgill1965generalgamma}. One of the most widely employed distribution for $RT$ data has been an exponentially modified Gaussian (or ex-Gaussian) distribution \cite{lacouture2008how, parris2013application, moret-tatay2018exgutils, balota1999word, heathcote1991analysis, navarro-pardo2013differences, gooch2012reaction, hervey2006reaction}, defined as the convolution of an exponential distribution with a Gaussian distribution. This distribution is characterized by three parameters, $\mu$, $\sigma$ and $\tau$, with $\mu$ and $\sigma$ characterizing the average and standard deviation of the Gaussian component and $\tau$ characterizing the decay rate of the exponential component. For $\sigma$ and $\tau$ greater than zero, the probability density function for the ex-Gaussian distribution is 
$$f(x) = \frac{1}{2\tau} \exp{\left(\frac{\sigma^2}{2\tau^2}-\frac{x-\mu}{\tau}\right)} \erfc\left(\frac{\sigma^2}{\tau}-({x-\mu}) \right) \enskip ,$$
where $\erfc$ is the complementary error function defined as
$$\erfc(x) = \frac{2}{\sqrt{\pi}}\int_{0}^{\infty} e^{-t^2} dt \enskip .$$
Researchers have attempted to specify the underlying process for $RT$ that leads to the ex-Gaussian distribution by attributing the exponential component to decision processes and the normal component to residual processes \cite{hohle1965inferred}. However, this rationale still remains unproven. Nonetheless, the ex-Gaussian distribution has been found to fit the $RT$ distribution better than gamma and log-normal distributions. 

\bibliographystyle{IEEEtran}
\bibliography{Trust_Group_Lib_References}

\end{document}